\documentclass[12pt]{article}

\usepackage{bbm,latexsym,graphicx}

\newcommand{\be}{\begin{equation}}
\newcommand{\ee}{\end{equation}}
\newcommand{\ba}{\begin{eqnarray}}
\newcommand{\ea}{\end{eqnarray}}
\newcommand{\no}{\nonumber\\}

\newcommand{\lesssim}{\:\mbox{\raisebox{-3pt}{$\stackrel%
{\displaystyle <}{\sim}$}}\:}

\newcommand{\mnu}{\mathcal{M}_\nu}

\begin{document}

\title{\normalsize \hfill UWThPh-2005-18 \\
\normalsize \hfill OCHA-PP-255 \\
\normalsize \hfill NIIG-DP-05-2 \\*[4mm] 
\LARGE $\mu$--$\tau$ antisymmetry and neutrino mass matrices}

\author{Walter Grimus\thanks{E-mail: walter.grimus@univie.ac.at} \\
\small Institut f\"ur Theoretische Physik, Universit\"at Wien \\
\small Boltzmanngasse 5, A--1090 Wien, Austria
\\*[3mm]
Satoru Kaneko\thanks{E-mail: satoru@phys.ocha.ac.jp} \\
\small Department of Physics, Ochanomizu University,
Tokyo 112-8610, Japan
\\*[3mm]
Lu\'{\i}s Lavoura\thanks{E-mail: balio@cftp.ist.utl.pt} \\
\small Universidade T\'ecnica de Lisboa
and Centro de F\'\i sica Te\'orica de Part\'\i culas \\
\small Instituto Superior T\'ecnico, 1049-001 Lisboa, Portugal
\\*[3mm]
Hideyuki Sawanaka\thanks{E-mail: hide@muse.sc.niigata-u.ac.jp} \\
\small Graduate School of Science and Technology, Niigata University,
\small 950-2181 Niigata, Japan
\\*[3mm]
Morimitsu Tanimoto\thanks{E-mail: tanimoto@muse.sc.niigata-u.ac.jp} \\
\small Department of Physics, Niigata University,
950-2181 Niigata, Japan
\\*[4mm]}

\date{25 October 2005}

\maketitle

\begin{abstract}
Using the seesaw mechanism and a discrete symmetry,
we construct a class of models for the neutrino mass matrix
where \emph{the inverse} of that matrix is the sum
of a $\mu$--$\tau$ antisymmetric background
and a perturbation.
We consider various possibilities for that perturbation.
The simplest possible perturbations lead to
four-parameter neutrino mass matrices
which are unable to fit the experimental data.
More complicated perturbations
give rise to viable six-parameter mass matrices;
we present detailed predictions of each of them.
\end{abstract}

\newpage

\section{Introduction}

The present experimental data on lepton mixing~\cite{tortola,fogli}
are compatible with maximal atmospheric neutrino mixing.
One possible theoretical explanation 
(for recent reviews see~\cite{altarelli})
for that maximal mixing,
if it indeed occurs,
is $\mu$--$\tau$ interchange symmetry~\cite{early,joshipura}.
In the basis where the charged-lepton mass matrix is diagonal, 
we denote the effective light-neutrino Majorana mass matrix by $\mnu$.
The diagonalization of $\mnu$ is performed as
\be
U^T \mnu\, U = \mathrm{diag} \left( m_1,\, m_2,\, m_3 \right),
\ee
where $m_{1,2,3}$ are the neutrino masses
and $U$ is the (unitary) lepton mixing matrix.
If $\mnu$ is $\mu$--$\tau$ symmetric,
i.e.~if $\mnu = M^\mathrm{(S)}$ with
\be
M^\mathrm{(S)} = \left( \begin{array}{ccc}
x & y & y \\ y & z & w \\ y & w & z
\end{array} \right),
\label{symmetric}
\ee
then atmospheric neutrino mixing is maximal,
i.e.~$\left| U_{\mu3} \right| = \left| U_{\tau3} \right|$.
Indeed,
the matrix~(\ref{symmetric}) has six physical parameters,
since only the phases of $z w^\ast$ and of $y^2 x^\ast z^\ast$ are physical.
The existence of only six physical parameters
in a theory for nine observables---the three neutrino masses $m_{1,2,3}$,
the three lepton mixing angles $\theta_{23,13,12}$,
one Dirac phase $\delta$,
and two Majorana phases $\Theta$ and $\Omega$---leads to
three predictions.\footnote{We defer the precise definition
of the mixing angles,
Dirac phase,
and Majorana phases to section~3.}
In this case the predictions are the following:
$\left| U_{\mu3} \right| = \left| U_{\tau3} \right|$
($\theta_{23} = \pi / 4$),
$U_{e3} = 0$
($\theta_{13} = 0$),
and the Dirac phase $\delta$ is meaningless
since $U$ has one vanishing matrix element.

The $\mu$--$\tau$ interchange symmetry,
represented by the matrix
\be
T = \left( \begin{array}{ccc}
1 & 0 & 0 \\ 0 & 0 & 1 \\ 0 & 1 & 0
\end{array} \right),
\ee
is a $\mathbbm{Z}_2$ symmetry,
and as such it has two possible eigenvalues:
$+1$ and $-1$.
The matrix $M^\mathrm{(S)}$ corresponds to the eigenvalue $+1$:
$T M^\mathrm{(S)} T = + M^\mathrm{(S)}$.
But there is also the possibility
that $\mnu$ is $\mu$--$\tau$ antisymmetric~\cite{aizawa},
i.e.~that $\mnu = M^\mathrm{(AS)}$,
where $T M^\mathrm{(AS)} T = - M^\mathrm{(AS)}$ or
\be
M^\mathrm{(AS)} = \left( \begin{array}{ccc}
0 & y & -y \\ y & z & 0 \\ -y & 0 & -z
\end{array} \right).
\label{antisymmetric}
\ee
This possibility seems attractive
since $M^\mathrm{(AS)}$ may be diagonalized as
\be
V^T M^\mathrm{(AS)}\, V =
{\rm diag} \left( i k,\, - i k,\, 0 \right),
\ee
where $k = \sqrt{2 \left| y \right|^2 + \left| z \right|^2}$
and the unitary matrix $V$ is
\be
V = \frac{1}{2 k} \left( \begin{array}{ccc}
2 y^\ast & 2 y^\ast & - 2 z \\
z^\ast + i k & z^\ast - i k & 2 y \\
z^\ast - i k & z^\ast + i k & 2 y
\end{array} \right).
\ee
Thus,
$\mnu = M^\mathrm{(AS)}$ leads to two degenerate massive neutrinos,
one massless neutrino,
maximal atmospheric neutrino mixing,
and maximal solar neutrino mixing
($\left| U_{e1} \right| = \left| U_{e2} \right|$
or $\theta_{12} = \pi / 4$).
These predictions are not so far from reality,
if the neutrino mass spectrum is inverted.
However, the exact equality $\mnu = M^\mathrm{(AS)}$ can be excluded,
since we know that all three neutrino masses are non-degenerate
and that solar neutrino mixing is large but not maximal.

In this paper we consider
various possible perturbations to $\mu$--$\tau$ antisymmetry.
More precisely, our starting point is 
\be\label{start}
\mnu^{-1} = M^\mathrm{(AS)} + M^\mathrm{(P)},
\ee
where $M^\mathrm{(P)}$ is a $\mu$--$\tau$ symmetric matrix, a
``perturbation'' to $M^\mathrm{(AS)}$.
Having $\mnu^{-1}$ instead of $\mnu$
in equation~(\ref{start}) is quite advantageous for model building;
this will become evident in section~2,
where we construct a class of models
based on equation~(\ref{start}).
Physically,
exchanging $\mnu^{-1}$ with $\mnu$
simply corresponds to complex-conjugating $U$
and trading the neutrino masses by their inverses.
In our class of models, the perturbation $M^\mathrm{(P)}$
is mostly free,
since it depends on the assumed scalar content of each model.
We proceed in the following sections
to analyze in detail the consequences of some perturbations.
We show analytically in sections 3--5 that
the simplest possible perturbations
lead to very constrained neutrino mass matrices,
which contradict the experimental data.
We proceed in section 6 to analyze numerically
some more complicated perturbations 
wherein $\mnu$ has six physical parameters.
We summarize our findings in section 7.

\section{A class of models}

Consider a $SU(2) \times U(1)$ gauge theory
with three scalar $SU(2)$ doublets $\phi_j$
($j = 1,2,3$).
Besides the left-handed lepton $SU(2)$ doublets $D_\alpha$
and the right-handed charged-lepton $SU(2)$ singlets $\alpha_R$
($\alpha = e, \mu, \tau$),
we introduce right-handed neutrinos $\nu_{\alpha R}$.
We assume the validity
of the lepton-number $U(1)$ symmetries $L_\alpha$
for all dimension-4 terms in the Lagrangian;
dimension-3 terms are allowed to softly break the $L_\alpha$.
We assume the existence of a symmetry $\mathbbm{Z}_2^{\rm (aux)}$
under which $\phi_1$,
$e_R$,
and all the $\nu_{\alpha R}$ change sign,
while all other fields remain invariant.
The Yukawa Lagrangian of the scalar doublets\footnote{We shall soon
introduce some scalar $SU(2)$ singlets into the theory,
which will have further Yukawa interactions.} is then
\ba
\mathcal{L}_{\mathrm{Y}\phi} &=& - \left(
y_1 \bar D_e \nu_{eR} +
y_2 \bar D_\mu \nu_{\mu R} +
y_2^\prime \bar D_\tau \nu_{\tau R}
\right) i \tau_2 \phi_1^\ast
- y_3 \bar D_e e_R \phi_1
\no & &
- \left( y_4 \bar D_\mu \mu_R + y_4^\prime \bar D_\tau \tau_R \right)
\phi_2
- \left( y_5 \bar D_\mu \mu_R + y_5^\prime \bar D_\tau \tau_R \right)
\phi_3
+ \rm{H.c.}
\ea
In the $\mu$--$\tau$ symmetric model
built by two of us a few years ago~\cite{Z2model},
a $\mu$--$\tau$ interchange symmetry
$\mathbbm{Z}_2^{\rm (tr)}$ was introduced,
under which $D_\mu \leftrightarrow D_\tau$,
$\mu_R \leftrightarrow \tau_R$,
$\nu_{\mu R} \leftrightarrow \nu_{\tau R}$,
and $\phi_3 \to - \phi_3$,
such that after two applications of $\mathbbm{Z}_2^{\rm (tr)}$
all fields transform into themselves.
Instead,
in the present model we employ a symmetry $\mathbbm{Z}_4$
under which
\be
\begin{array}{rclcrclcrcl}
D_e &\to& i D_e,
& \ &
D_\mu &\to& i D_\tau,
& \ &
D_\tau &\to& i D_\mu,
\\
e_R &\to& i e_R,
& \ &
\mu_R &\to& i \tau_R,
& \ &
\tau_R &\to& i \mu_R,
\\
\nu_{eR} &\to& i \nu_{eR},
& \ &
\nu_{\mu R} &\to& i \nu_{\tau R},
& \ &
\nu_{\tau R} &\to& i \nu_{\mu R},
\\
\phi_3 &\to& - \phi_3.
\end{array}
\ee
Notice that all lepton fields change sign
after two applications of this $\mathbbm{Z}_4$.
Just as $\mathbbm{Z}_2^{\rm (tr)}$ in~\cite{Z2model},
the present $\mathbbm{Z}_4$ enforces $y_2^\prime = y_2$,
$y_4^\prime = y_4$,
and $y_5^\prime = - y_5$.
The neutrino Dirac mass matrix $M_D$
and the charged-lepton mass matrix $M_\ell$ are then
\ba
M_D &=& {\rm diag} \left( a,\, b,\, b \right),
\label{MD} \\
M_\ell &=& {\rm diag} \left( e^{i \varphi_e} m_e,\,
e^{i \varphi_\mu} m_\mu,\, e^{i \varphi_\tau} m_\tau \right),
\label{Mell}
\ea
respectively,
where $a = y_1^\ast v_1$,
$b = y_2^\ast v_1$,
$m_e = \left| y_3 v_1 \right|$,
$m_\mu = \left| y_4 v_2 + y_5 v_3 \right|$,
and $m_\tau = \left| y_4 v_2 - y_5 v_3 \right|$,
the $v_j$ being the vacuum expectation values (VEVs)
of the lower components of the $\phi_j$.

Now consider the Majorana mass terms of the right-handed neutrinos,
\be
{\cal L}_{\rm Majorana} =
\frac{1}{2}
\left( \nu_{eR}^T,\, \nu_{\mu R}^T,\, \nu_{\tau R}^T \right)
C^{-1} M_R^\ast \left( \begin{array}{c}
\nu_{eR} \\ \nu_{\mu R} \\ \nu_{\tau R}
\end{array} \right)
+ {\rm H.c.}
\ee
They have dimension 3 and are,
therefore,
allowed to break the lepton-number symmetries $L_\alpha$.
Because of the symmetry $\mathbbm{Z}_4$,
which is \emph{not} allowed to be broken softly,
one has
\be
M_R = M^\mathrm{(AS)}.
\label{MR}
\ee
The seesaw mechanism~\cite{seesaw,seesaw1} prescribes
\be
\mnu = - M_D^T M_R^{-1} M_D,
\label{seesaw}
\ee
hence
\be
\mnu^{-1} = - M_D^{-1} M_R {M_D^T}^{-1}.
\ee
It follows from equations~(\ref{MD}) and~(\ref{MR}) that
$\mnu^{-1}$ is $\mu$--$\tau$ antisymmetric,
just as $M_R$.

The model as it now stands is not realistic for many reasons,
in particular because $M_R$ in equation~(\ref{MR}) is not invertible
and therefore the seesaw formula~(\ref{seesaw}) cannot apply.
One may correct for this by adding to the model
one or more scalar $SU(2) \times U(1)$ invariants
with non-vanishing lepton numbers.
The simplest possibilities are the following:
\begin{enumerate}
\item A complex scalar $\chi_{ee}$ with $L_e = -2$
and $L_\mu = L_\tau = 0$,
which changes sign under $\mathbbm{Z}_4$.
Its Yukawa coupling
$\chi_{ee} \nu_{eR}^T C^{-1} \nu_{eR} + {\rm H.c.}$
generates a non-vanishing $\left( M_R \right)_{ee}$
upon $\chi_{ee}$ acquiring a VEV.
One then has
\be
\mathrm{case\ 1}: \quad \mnu^{-1} = \left( \begin{array}{ccc}
x & y & -y \\ y & z & 0 \\ -y & 0 & -z
\end{array} \right).
\label{chiee}
\ee
\item A complex scalar $\chi_{\mu\tau}$
with $L_e = 0$ and $L_\mu = L_\tau = -1$,
which changes sign under $\mathbbm{Z}_4$.
Its Yukawa coupling
$\chi_{\mu\tau} \nu_{\mu R}^T C^{-1} \nu_{\tau R} + {\rm H.c.}$
generates a non-vanishing $\left( M_R \right)_{\mu \tau}$
upon $\chi_{\mu\tau}$ acquiring a VEV.
One then has
\be
\mathrm{case\ 2}: \quad \mnu^{-1} = \left( \begin{array}{ccc}
0 & y & -y \\ y & z & w \\ -y & w & -z
\end{array} \right).
\label{chimutau}
\ee
\item Two complex scalars $\chi_{e\mu}$ and $\chi_{e\tau}$,
$\chi_{e\mu}$ having $L_e = L_\mu = -1$ and $L_\tau = 0$
while $\chi_{e\tau}$ has $L_e = L_\tau = -1$ and $L_\mu = 0$.
Under $\mathbbm{Z}_4$,
$\chi_{e\mu} \leftrightarrow - \chi_{e\tau}$.
The Yukawa couplings
$\nu_{eR}^T C^{-1} \left( \chi_{e\mu} \nu_{\mu R}
+ \chi_{e\tau} \nu_{\tau R} \right) + {\rm H.c.}$
lead to
$\left( M_R \right)_{e\mu} \neq - \left( M_R \right)_{e\tau}$
upon $\chi_{e\mu}$ and $\chi_{e\tau}$ acquiring VEVs.
One thus obtains
\be
\mathrm{case\ 3}: \quad \mnu^{-1} = \left( \begin{array}{ccc}
0 & y & -y^\prime \\ y & z & 0 \\ -y^\prime & 0 & -z
\end{array} \right).
\label{chiemu}
\ee
\item Two complex scalars $\chi_{\mu\mu}$ and $\chi_{\tau\tau}$,
$\chi_{\mu\mu}$ having $L_e = L_\tau = 0$ and $L_\mu = -2$
while $\chi_{\tau\tau}$ has $L_e = L_\mu = 0$ and $L_\tau = -2$.
Under $\mathbbm{Z}_4$,
$\chi_{\mu\mu} \leftrightarrow - \chi_{\tau\tau}$.
The Yukawa couplings
$\chi_{\mu\mu} \nu_{\mu R}^T C^{-1}\nu_{\mu R}
+ \chi_{\tau\tau} \nu_{\tau R}^T C^{-1} \nu_{\tau R} + {\rm H.c.}$
lead to
$\left( M_R \right)_{\mu\mu} \neq - \left( M_R \right)_{\tau\tau}$.
One then has
\be
\mathrm{case\ 4}: \quad \mnu^{-1} = \left( \begin{array}{ccc}
0 & y & -y \\ y & z & 0 \\ -y & 0 & -z^\prime
\end{array} \right).
\label{chimumu}
\ee
\end{enumerate}
One may put together any two of the possibilities 1--4
and obtain a neutrino mass matrix
with more degrees of freedom---and
correspondingly less predictive power.
One thus obtains six more possibilities:
\ba
\mathrm{case\ 5}: & & \mnu^{-1} = \left( \begin{array}{ccc}
x & y & -y \\ y & z & w \\ -y & w & -z
\end{array} \right);
\label{five} \\
\mathrm{case\ 6}: & & \mnu^{-1} = \left( \begin{array}{ccc}
x & y & -y^\prime \\ y & z & 0 \\ -y^\prime & 0 & -z
\end{array} \right);
\label{six} \\
\mathrm{case\ 7}: & & \mnu^{-1} = \left( \begin{array}{ccc}
x & y & -y \\ y & z & 0 \\ -y & 0 & -z^\prime
\end{array} \right);
\label{seven} \\
\mathrm{case\ 8}: & & \mnu^{-1} = \left( \begin{array}{ccc}
0 & y & -y^\prime \\ y & z & w \\ -y^\prime & w & -z
\end{array} \right);
\label{eight} \\
\mathrm{case\ 9}: & & \mnu^{-1} = \left( \begin{array}{ccc}
0 & y & -y \\ y & z & w \\ -y & w & -z^\prime
\end{array} \right);
\label{nine} \\
\mathrm{case\ 10}: & & \mnu^{-1} = \left( \begin{array}{ccc}
0 & y & -y^\prime \\ y & z & 0 \\ -y^\prime & 0 & -z^\prime
\end{array} \right).
\label{ten}
\ea
It is the purpose of the rest of this paper to study,
both analytically and numerically,
whether cases 1--10 are viable or not,
and to find out their predictions.

The neutrino mass matrix $\mnu$ may be rephased:
$\mnu \to X \mnu X$,
where $X$ is a diagonal unitary matrix.
Because of this possibility,
in general only three phases in the symmetric $\mnu$
are physically meaningful.
Using this rephasing freedom,
one finds that cases 1--4 have four physical parameters each,
case 10 has five physical parameters,
and the remaining cases 5--9 have six physical parameters each.

\section{Cases 3, 4, and 10 are not viable}

These three cases contradict experiment
since in all three of them the conditions
\ba
\left( \mnu^{-1} \right)_{ee} &=& 0,
\label{ee}
\\
\left( \mnu^{-1} \right)_{\mu\tau} &=& 0
\label{mutau}
\ea
simultaneously hold~\cite{lavoura}.
Using
\be
\mnu^{-1} =
U\, \mathrm{diag} \left( m_1^{-1},\, m_2^{-1},\, m_3^{-1} \right) U^T,
\label{mnu-1}
\ee
we find that the system
of the two conditions~(\ref{ee}) and (\ref{mutau}) leads to
\ba
\frac{m_1}{m_3} &=&
\frac{U_{e1} U_{\tau2} U_{\tau3}^\ast - U_{e2} U_{\mu1} U_{\mu3}^\ast}
{U_{e2} U_{\tau3} U_{\tau1}^\ast - U_{e3} U_{\mu2} U_{\mu1}^\ast},
\label{m13} \\
\frac{m_2}{m_3} &=&
\frac{U_{e1} U_{\tau2} U_{\tau3}^\ast - U_{e2} U_{\mu1} U_{\mu3}^\ast}
{U_{e3} U_{\tau1} U_{\tau2}^\ast - U_{e1} U_{\mu3} U_{\mu2}^\ast}.
\label{m23}
\ea
We use the standard parametrization
\be
U = P_F\, \hat U P_M,
\ee
where
\be
\hat U = \left( \begin{array}{ccc}
c_{13} c_{12} & c_{13} s_{12} & s_{13} e^{- i \delta} \\
- c_{23} s_{12} - s_{23} s_{13} c_{12} e^{i \delta} &
c_{23} c_{12} - s_{23} s_{13} s_{12} e^{i \delta} &
s_{23} c_{13} \\
- s_{23} s_{12} + c_{23} s_{13} c_{12} e^{i \delta} &
s_{23} c_{12} + c_{23} s_{13} s_{12} e^{i \delta} &
- c_{23} c_{13}
\end{array} \right),
\label{formU}
\ee
\be
P_F = \mathrm{diag} \left(  
e^{i \vartheta_e},\, e^{i \vartheta_\mu},\, e^{i \vartheta_\tau} \right),
\label{PF}
\ee
\be
P_M = \mathrm{diag} \left(
e^{i \Theta / 2},\, 1,\, e^{i \Omega / 2} \right).
\label{reU}
\ee
In equation~(\ref{formU}),
$s_k = \sin{\theta_k}$ and $c_k = \cos{\theta_k}$
for $k = 23, 13, 12$.
The phases $\vartheta_\alpha$ in equation~(\ref{PF}) are unobservable.
Using the parametrization above,
equations~(\ref{m13}) and~(\ref{m23}) translate into
\ba
\frac{m_1}{m_3} &=& e^{i \left( \Theta - \Omega \right)}\,
\frac
{\left( 1 - \left| \epsilon \right|^2 \right)
\left[ c_{23} s_{23} \left( s_{12}^2 - c_{12}^2 \right)
+ c_{12} s_{12} \left( s_{23}^2 - c_{23}^2 \right) \epsilon \right]}
{c_{23} s_{23} s_{12}^2 \left( 1 - 2 \left| \epsilon \right|^2 \right)
+ c_{23} s_{23} c_{12}^2 {\epsilon^\ast}^2
+ c_{12} s_{12} \left( c_{23}^2 - s_{23}^2 \right)
\left| \epsilon \right|^2 \epsilon^\ast},
\label{k1} \\
\frac{m_2}{m_3} &=& e^{- i \Omega}\,
\frac
{\left( 1 - \left| \epsilon \right|^2 \right)
\left[ c_{23} s_{23} \left( s_{12}^2 - c_{12}^2 \right)
+ c_{12} s_{12} \left( s_{23}^2 - c_{23}^2 \right) \epsilon \right]}
{- c_{23} s_{23} c_{12}^2 \left( 1 - 2 \left| \epsilon \right|^2 \right)
- c_{23} s_{23} s_{12}^2 {\epsilon^\ast}^2
+ c_{12} s_{12} \left( c_{23}^2 - s_{23}^2 \right)
\left| \epsilon \right|^2 \epsilon^\ast},
\label{k2}
\ea
where we have defined $\epsilon \equiv s_{13} e^{i \delta}$.
Experimentally~\cite{tortola},
both $\left| \epsilon \right|$ and $\left| c_{23}^2 - s_{23}^2 \right|$
are at most of order $0.2$;
neglecting terms of order
$\epsilon^2$,
$\left( c_{23}^2 - s_{23}^2 \right)^2$,
and $\epsilon \left( c_{23}^2 - s_{23}^2 \right)$
in equations~(\ref{k1}) and~(\ref{k2}),
one obtains
\ba
\frac{m_1}{m_3} &\approx& 1 - \frac{c_{12}^2}{s_{12}^2},
\\
\frac{m_2}{m_3} &\approx& 1 - \frac{s_{12}^2}{c_{12}^2}.
\ea
Using the experimental result~\cite{tortola}
$\tan^2{\theta_{12}} \approx 0.43$,
one obtains
\be
\frac{m_2^2 - m_1^2}{m_3^2 - m_1^2} \approx 1.9.
\label{uytre}
\ee
Although the number in the right-hand side of equation~(\ref{uytre})
is rather sensitive to the precise value of  $\tan^2{\theta_{12}}$,
its order of magnitude does not change.
It is in manifest contradiction with experiment,
which gives
\be
\left| \frac{m_2^2 - m_1^2}{m_3^2 - m_1^2} \right| \approx 0.037.
\ee
Thus,
cases 3,
4,
and 10 are excluded.

\section{Case 1 is not viable}

Define the orthogonal matrix
\be
R_1 = \left( \begin{array}{ccc}
1 & 0 & 0 \\ 0 & r & r \\ 0 & -r & r
\end{array} \right),
\quad \mathrm{where} \quad r = \frac{1}{\sqrt{2}}.
\ee
Then,
with the $\mnu^{-1}$ of case 1,
in equation~(\ref{chiee}),
one has
\be
R_1^T \mnu^{-1} R_1 = \left( \begin{array}{ccc}
x & \sqrt{2} y & 0 \\ \sqrt{2} y & 0 & z \\ 0 & z & 0
\end{array} \right).
\label{utyro}
\ee
This is a mass matrix of the Fritzsch type,
which may be diagonalized in a standard fashion~\cite{fritzsch}.
The three phases in the matrix~(\ref{utyro}) may be discarded
by using a diagonal unitary matrix $P_1$:
\be
P_1 R_1^T \mnu^{-1} R_1 P_1 = \left( \begin{array}{ccc}
\left| x \right| & \sqrt{2} \left| y \right| & 0 \\
\sqrt{2} \left| y \right| & 0 & \left| z \right| \\
0 & \left| z \right| & 0
\end{array} \right).
\label{m}
\ee
This matrix has eigenvalues $\lambda_1$,
$- \lambda_2$,
and $\lambda_3$,
with $\lambda_1 > \lambda_2 > \lambda_3 > 0$.
If we define the diagonal unitary matrix
\be
P_2 = \mathrm{diag} \left( 1,\, i,\, 1 \right),
\ee
then there is an orthogonal matrix $R_2$ such that
\be
P_2 R_2^T P_1 R_1^T \mnu^{-1} R_1 P_1 R_2 P_2 =
\mathrm{diag} \left( \lambda_3,\, \lambda_2,\, \lambda_1 \right).
\ee
The standard diagonalization of the Fritzsch mass matrix yields 
\ba
\left( R_2 \right)_{11} &=&
\sqrt{\frac{\lambda_3 \left( \lambda_1 + \lambda_3 \right)
\left( \lambda_2 - \lambda_3 \right)}
{\left( \lambda_1 - \lambda_3 \right)
\left( \lambda_2 + \lambda_3 \right)
\left( \lambda_1 - \lambda_2 + \lambda_3 \right)}},
\\
\left( R_2 \right)_{12} &=&
\sqrt{\frac{\lambda_2 \left( \lambda_1 - \lambda_2 \right)
\left( \lambda_2 - \lambda_3 \right)}
{\left( \lambda_1 + \lambda_2 \right)
\left( \lambda_2 + \lambda_3 \right)
\left( \lambda_1 - \lambda_2 + \lambda_3 \right)}},
\\
\left( R_2 \right)_{13} &=&
\sqrt{\frac{\lambda_1 \left( \lambda_1 - \lambda_2 \right)
\left( \lambda_1 + \lambda_3 \right)}
{\left( \lambda_1 + \lambda_2 \right)
\left( \lambda_1 - \lambda_3 \right)
\left( \lambda_1 - \lambda_2 + \lambda_3 \right)}}.
\ea
Since
\be
U^\dagger \mnu^{-1} U^\ast = \mathrm{diag} \left(
m_1^{-1},\, m_2^{-1},\, m_3^{-1} \right),
\label{m-1}
\ee
we see that there are two possibilities:
\begin{enumerate}
\item If the neutrino mass spectrum is normal,
$m_1 < m_2 < m_3$,
then $\lambda_j = m_j^{-1}$ for $j = 1, 2, 3$ and
\be
\left| \frac{U_{e2}}{U_{e1}} \right|^2 =
\left[ \frac{\left( R_2 \right)_{12}}{\left( R_2 \right)_{13}} \right]^2
= \frac{m_2^{-1} \left( m_2^{-1} - m_3^{-1} \right)
\left( m_1^{-1} - m_3^{-1} \right)}
{m_1^{-1} \left( m_1^{-1} + m_3^{-1} \right)
\left( m_2^{-1} + m_3^{-1} \right)}.
\label{case1}
\ee
\item If the neutrino mass spectrum is inverted,
$m_3 < m_1 < m_2$,
then $\lambda_1 = m_3^{-1}$,
$\lambda_2 = m_1^{-1}$,
and $\lambda_3 = m_2^{-1}$.
One then has
\be
\left| \frac{U_{e2}}{U_{e1}} \right|^2 =
\left[ \frac{\left( R_2 \right)_{11}}{\left( R_2 \right)_{12}} \right]^2
= \frac{m_2^{-1} \left( m_3^{-1} + m_2^{-1} \right)
\left( m_3^{-1} + m_1^{-1} \right)}
{m_1^{-1} \left( m_3^{-1} - m_1^{-1} \right)
\left( m_3^{-1} - m_2^{-1} \right)}.
\label{case2}
\ee
\end{enumerate}
The inverted case is readily excluded.
With $m_2^2 - m_1^2 = 8.1 \times 10^{-5}\, \mathrm{eV}^2$
and $\left| m_1^2 - m_3^2 \right| = 2.2 \times 10^{-3}\, \mathrm{eV}^2$,
one finds that equation~(\ref{case2}) yields
$\left| U_{e2} \left/ U_{e1} \right. \right|^2 > 0.98$ for all $m_1$,
which is far above the best-fit value
$\left| U_{e2} \left/ U_{e1} \right. \right|^2 = 0.43$.
As for equation~(\ref{case1}),
with the above values for the mass-squared differences, 
it always yields $\left| U_{e2} \left/ U_{e1} \right. \right|^2 < 0.28$.
This value translates into $s_{12}^2 < 0.22$.
Reporting here for further use 
the $3\sigma$ values derived in~\cite{tortola},
\ba
& & 7.2 \times 10^{-5}\, \mathrm{eV}^2
< m_2^2 - m_1^2 <
9.1 \times 10^{-5}\, \mathrm{eV}^2,
\label{21} \\
& & 1.4 \times 10^{-3}\, \mathrm{eV}^2
< \left| m_1^2 - m_3^2 \right| <
3.3 \times 10^{-3}\, \mathrm{eV}^2,
\label{31} \\
& & 0.23 < \sin^2{\theta_{12}} < 0.38,
\label{sol} \\
& & 0.34 < \sin^2{\theta_{23}} < 0.66,
\label{atm} \\
& & \sin^2{\theta_{13}} < 0.047,
\label{ue3}
\ea
we see that $0.22$ is just outside
the $3\sigma$ range of $\sin^2{\theta_{12}}$.
However,
the upper bound on $\left| U_{e2} \left/ U_{e1} \right. \right|^2$
of equation~(\ref{case1})
is sensitive to the values of the mass-squared differences.
Taking the lower limit
$m_2^2 - m_1^2 = 7.2 \times 10^{-5}\, \mathrm{eV}^2$
from inequality~(\ref{21}) and the upper limit 
$m_3^2 - m_1^2 = 3.3 \times 10^{-3}\, \mathrm{eV}^2$
from inequality~(\ref{31}),
the upper bound on 
$\left| U_{e2} \left/ U_{e1} \right. \right|^2$ is $0.35$,
or $s_{12}^2 < 0.26$,
which is inside the $3\sigma$ range of inequality~(\ref{sol}).
In that case,
a simultaneous consideration of 
$s_{13}^2 = \left[ \left( R_2 \right)_{11} \right]^2$
leads to $s_{13}^2 = 0.29$,
for the $m_1$ where the maximum of equation~(\ref{case1}) is reached.
In view of the inequality~(\ref{ue3}),
this value is clearly much too large, hence
case 1 is excluded.

\section{Case 2 is not viable}
\label{sec:case2}

Case 2 is defined by
\ba
\left( \mnu^{-1} \right)_{ee} &=& 0, \label{one}
\\
\left( \mnu^{-1} \right)_{e\mu} + \left( \mnu^{-1} \right)_{e\tau}
&=& 0, \label{two}
\\
\left( \mnu^{-1} \right)_{\mu\mu} + \left( \mnu^{-1} \right)_{\tau\tau}
&=& 0. \label{three}
\ea
In the following,
we shall use the quantities
\be
\mu_1 \equiv m_1 e^{- i \Theta}, \quad
\mu_2 \equiv m_2, \quad
\mu_3 \equiv m_3 e^{- i \Omega}.
\label{Mpi}
\ee
The matrix  elements $\left( \mnu^{-1} \right)_{\alpha\beta}$
are then given as
\be
(\mnu^{-1})_{\alpha\beta} =
e^{i \left( \vartheta_\alpha + \vartheta_\beta \right)}\,
\sum^3_{j=1}\, \frac{\hat U_{\alpha j}\hat U_{\beta j}}{\mu_j},
\ee
with $\hat U$ in equation~(\ref{formU}).
We shall moreover use
\be
\beta \equiv \vartheta_\tau - \vartheta_\mu,
\label{beta}
\ee
$\epsilon \equiv s_{13} \exp{\left( i \delta \right)}$,
and $\nu \equiv \cos{2 \theta_{23}}$.
We also introduce 
the small parameter $\lambda = 0.22$
which is useful since,
experimentally,
both $\left| \epsilon \right|$ and $\left| \nu \right|$
have upper bounds of order $\lambda$.
Moreover,
$\left. \left( m_2^2 - m_1^2 \right) \right/
\left| m_1^2 - m_3^2 \right| \approx 0.037$
is of order $\lambda^2$.

Equation~(\ref{one}) yields
\be
\frac{\left( \hat U_{e1} \right)^2}{\mu_1}
+ \frac{\left( \hat U_{e2} \right)^2}{\mu_2}
+ \frac{\left( \hat U_{e3} \right)^2}{\mu_3} = 0,
\ee
or
\be
\mu_3 = - \frac{{\epsilon^\ast}^2}
{1 - \left| \epsilon \right|^2}
\left( \frac{c_{12}^2}{\mu_1} + \frac{s_{12}^2}{\mu_2} \right)^{-1}.
\label{one'}
\ee
Taking into account that $m_2 > m_1$ and $c_{12} > s_{12}$,
equation~(\ref{one'}) leads to the inequality 
\be
\frac{m_3}{m_1} < \frac{\left| \epsilon \right|^2}
{1 - \left| \epsilon \right|^2}\,
\frac{1}{c_{12}^2 - s_{12}^2},
\ee
hence the neutrino mass spectrum in case 2 must be \emph{inverted}.
The quantity $c_{12}^2 - s_{12}^2$ is smaller than 1:
with the best-fit value $s_{12}^2 = 0.30$,
it is $0.40$;
at the upper edge of the $3 \sigma$ range for $s_{12}^2$,
it is only $0.24$,
i.e.~of order $\lambda$.
Thus,
it is appropriate to admit $m_3/m_1$
to be of order $\lambda$,
instead of $\lambda^2$.
As for $m_3/m_2$,
it is of the same order as $m_3/m_1$,
since
\be
\frac{m_3}{m_2} = \frac{m_3}{\sqrt{m_1^2 + \Delta m^2_\odot}}
\ee
and $m_1^2$ is of the order of the atmospheric mass-squared difference
when $m_3 / m_1$ is very small.
We obtain a picture in which
$m_1 \approx m_2 \approx \sqrt{\Delta m^2_\mathrm{atm}}$,
while $m_3 / m_1 \lesssim\, 0.2$ is small.

Equation~(\ref{two}) gives,
after using equation~(\ref{one'}),
\be
s_{23} - e^{i\beta} c_{23}
+ \zeta \left( c_{23} + e^{i\beta} s_{23} \right) = 0,
\label{1+2}
\ee
where
\be
\zeta \equiv \epsilon^\ast\,
\frac{\mu_2 - \mu_1}{c_{12}^2 \mu_2 + s_{12}^2 \mu_1}\, 
c_{12} s_{12}.
\label{zeta}
\ee

Next we address equation~(\ref{three}).
Defining
\ba
z &\equiv& c_{23}^2 e^{i\beta} + s_{23}^2 e^{-i\beta}
\no &=& \cos{\beta} + i \nu \sin{\beta},
\ea
we find that equation~(\ref{three}) may be written as
\be
az + bz^\ast + c = 0,
\label{eqz}
\ee
where
\ba
a &=& 1 - 2 \left| \epsilon \right|^2, \\
b &=& - {\epsilon^\ast}^2\,
\frac{s_{12}^2 \mu_2 + c_{12}^2 \mu_1}{c_{12}^2 \mu_2 + s_{12}^2 \mu_1},
\\ 
c &=& 2 i \left| \epsilon \right|^2 \sqrt{1 - \nu^2}\
\zeta \sin \beta,
\ea
where we have once again used equation~(\ref{one'})
and the definition~(\ref{zeta}).
The solution of equation~(\ref{eqz}) for $z$ is
\be
z = \frac{b c^\ast - a^\ast c}{\left| a \right|^2 - \left| b \right|^2}.
\label{zsolution}
\ee

An upper limit on the absolute value of $\zeta$ is 
\ba
\left| \zeta \right| &\leq&
\left| \epsilon \right| 
\frac{m_2 + m_1}{c_{12}^2 m_2 - s_{12}^2 m_1}\, c_{12} s_{12}
\no &=&
\left| \epsilon \right| \frac{2 \left( m_2 + m_1 \right)}
{\left( c_{12}^2 - s_{12}^2 \right) \left( m_1 + m_2 \right)
+ m_2 - m_1}\,
c_{12} s_{12}
\no &<&
\left| \epsilon \right| \tan{2 \theta_{12}}.
\ea
The $3\sigma$ limits $\left| \epsilon \right| < 0.22$
and $\tan{2 \theta_{12}} < 4.04$,
cf.~inequalities~(\ref{sol}) and (\ref{ue3}),
lead to $\left| \zeta \right| < 0.89$;
at the $2\sigma$ level---see~\cite{tortola}---this bound
is already down to $\left| \zeta \right| < 0.50$.

Let us now use simple approximations to show that case 2
is not viable.
Since $a \approx 1$ and $\left| b \right| \! \lesssim \lambda^2$,
we conclude from equation~(\ref{zsolution}) that $z \simeq - c$:
\be
\cos \beta + i \nu \sin \beta \simeq
- 2 i \left| \epsilon \right|^2 \sqrt{1 - \nu^2}\
\zeta \sin \beta.
\label{this}
\ee
The right-hand side of equation~(\ref{this})
being at most of order $\lambda^2$,
one must have
\be
e^{i\beta} \simeq \eta i\ (\eta = \pm 1),
\quad 
\nu \simeq 0,
\ee
up to corrections of order $\lambda^2$.
Therefore,
\be
c_{23} + e^{i\beta} s_{23} \simeq \frac{1 + \eta i}{\sqrt{2}}, 
\quad
s_{23} - e^{i\beta} c_{23} \simeq \frac{1 - \eta i}{\sqrt{2}}.
\label{finaltwo}
\ee
Introducing the approximations~(\ref{finaltwo}) into equation~(\ref{1+2}),
one finds
\be
\zeta \simeq \frac{- 1 + i \eta}{1 + i \eta} = i \eta,
\label{A}
\ee
i.e.~$\zeta = \pm i$.
But we know that at $3 \sigma$ level
$\left| \zeta \right|$ is at most $0.89$.
Hence case 2 is ruled out.

\section{Cases 5--9 are viable}

Case 5 is defined by conditions~(\ref{two}) and~(\ref{three}).
Using the phase $\beta$ of the definition~(\ref{beta}),
they read
\ba
\frac{\hat U_{e1} \left( \hat U_{\mu 1} +
e^{i \beta} \hat U_{\tau 1} \right)}{\mu_1}
+ \frac{\hat U_{e2} \left( \hat U_{\mu 2} +
e^{i \beta} \hat U_{\tau 2} \right)}{\mu_2}
+ \frac{\hat U_{e3} \left( \hat U_{\mu 3} +
e^{i \beta} \hat U_{\tau 3} \right)}{\mu_3} &=& 0,
\label{c5cond1} \\
\frac{\left[ \left( \hat U_{\mu 1} \right)^2 +
e^{2 i \beta} \left( \hat U_{\tau 1} \right)^2 \right]}{\mu_1}
+ \frac{\left[ \left( \hat U_{\mu 2} \right)^2 +
e^{2 i \beta} \left( \hat U_{\tau 2} \right)^2 \right]}{\mu_2}
+ \frac{\left[ \left( \hat U_{\mu 3} \right)^2 +
e^{2 i \beta} \left( \hat U_{\tau 3} \right)^2 \right]}{\mu_3}
&=& 0. \hspace*{5mm}
\label{c5cond2}
\ea
These two equations lead to
\ba
\frac{\mu_1}{\mu_3} &=&
\frac{\hat U_{e2}
\left( \hat U_{\mu2} + e^{i \beta} \hat U_{\tau2} \right)
\left( \hat U_{\mu1}^2 + e^{2 i \beta} \hat U_{\tau1}^2 \right)
- \hat U_{e1}
\left( \hat U_{\mu1} + e^{i \beta} \hat U_{\tau1} \right)
\left( \hat U_{\mu2}^2 + e^{2 i \beta} \hat U_{\tau2}^2 \right)}
{\hat U_{e3}
\left( \hat U_{\mu3} + e^{i \beta} \hat U_{\tau3} \right)
\left( \hat U_{\mu2}^2 + e^{2 i \beta} \hat U_{\tau2}^2 \right)
- \hat U_{e2}
\left( \hat U_{\mu2} + e^{i \beta} \hat U_{\tau2} \right)
\left( \hat U_{\mu3}^2 + e^{2 i \beta} \hat U_{\tau3}^2 \right)},
\label{ratio5} \\
\frac{\mu_2}{\mu_3} &=& 
\frac{\hat U_{e2}
\left( \hat U_{\mu2} + e^{i \beta} \hat U_{\tau2} \right)
\left( \hat U_{\mu1}^2 + e^{2 i \beta} \hat U_{\tau 1}^2 \right)
- \hat U_{e1}
\left( \hat U_{\mu1} + e^{i \beta} \hat U_{\tau1} \right)
\left( \hat U_{\mu2}^2 + e^{2 i \beta} \hat U_{\tau2}^2 \right)}
{\hat U_{e1}
\left( \hat U_{\mu1} + e^{i \beta} \hat U_{\tau1} \right)
\left( \hat U_{\mu3}^2 + e^{2 i \beta} \hat U_{\tau3}^2 \right)
- \hat U_{e3}
\left( \hat U_{\mu3} + e^{i \beta} \hat U_{\tau3} \right)
\left( \hat U_{\mu1}^2 + e^{2 i \beta} \hat U_{\tau1}^2 \right)}.
\hspace*{5mm}
\label{ratio6}
\ea

In each of cases 6--9 there is a set of two conditions of the type
$\left( \mnu^{-1} \right)_{\alpha\beta} = 0$
and $\left( \mnu^{-1} \right)_{\gamma\delta}
+ \left( \mnu^{-1} \right)_{\rho\sigma} = 0$.
By solving the system of these two equations one obtains
\ba
\frac{\mu_1}{\mu_3} & = & 
\frac{\left( \hat U_{\gamma2} \hat U_{\delta2}
+ e^{i \varphi} \hat U_{\rho2} \hat U_{\sigma2} \right)
\hat U_{\alpha1} \hat U_{\beta1}
- \left( \hat U_{\gamma1} \hat U_{\delta1}
+ e^{i \varphi} \hat U_{\rho1} \hat U_{\sigma1} \right)
\hat U_{\alpha2} \hat U_{\beta2}}
{\left( \hat U_{\gamma3} \hat U_{\delta3}
+ e^{i \varphi} \hat U_{\rho3} \hat U_{\sigma3} \right)
\hat U_{\alpha2} \hat U_{\beta2}
- \left( \hat U_{\gamma2}\hat U_{\delta2}
+ e^{i \varphi} \hat U_{\rho2} \hat U_{\sigma2} \right)
\hat U_{\alpha3} \hat U_{\beta3}},
\label{rati} \\
\frac{\mu_2}{\mu_3} &=&
\frac{\left( \hat U_{\gamma2} \hat U_{\delta2}
+ e^{i \varphi} \hat U_{\rho2} \hat U_{\sigma2} \right)
\hat U_{\alpha1} \hat U_{\beta1}
- \left( \hat U_{\gamma1} \hat U_{\delta1}
+ e^{i \varphi} \hat U_{\rho1} \hat U_{\sigma1} \right)
\hat U_{\alpha2} \hat U_{\beta2}}
{\left( \hat U_{\gamma1} \hat U_{\delta1}
+ e^{i \varphi} \hat U_{\rho1} \hat U_{\sigma1} \right)
\hat U_{\alpha3} \hat U_{\beta3}
- \left( \hat U_{\gamma3} \hat U_{\delta3}
+ e^{i \varphi} \hat U_{\rho3} \hat U_{\sigma3} \right)
\hat U_{\alpha1} \hat  U_{\beta1}},
\label{ratio}
\ea
where $\varphi \equiv \vartheta_\rho + \vartheta_\sigma
- \vartheta_\gamma - \vartheta_\delta$.
The unphysical phase $\varphi$ arises because
the relative phase in the equality
$\left( \mnu^{-1} \right)_{\gamma\delta}
= - \left( \mnu^{-1} \right)_{\rho\sigma}$
is arbitrary due to the rephasing freedom.

The right-hand sides of equations~(\ref{ratio5})--(\ref{ratio})
are functions of the observables $\theta_{12}$,
$\theta_{23}$,
$\theta_{13}$,
and $\delta$,
plus an unphysical phase---$\beta$
in equations~(\ref{ratio5}) and~(\ref{ratio6}),
$\varphi$ in equations~(\ref{rati}) and~(\ref{ratio}).

\begin{figure}
\begin{center}
\includegraphics[scale=.5]{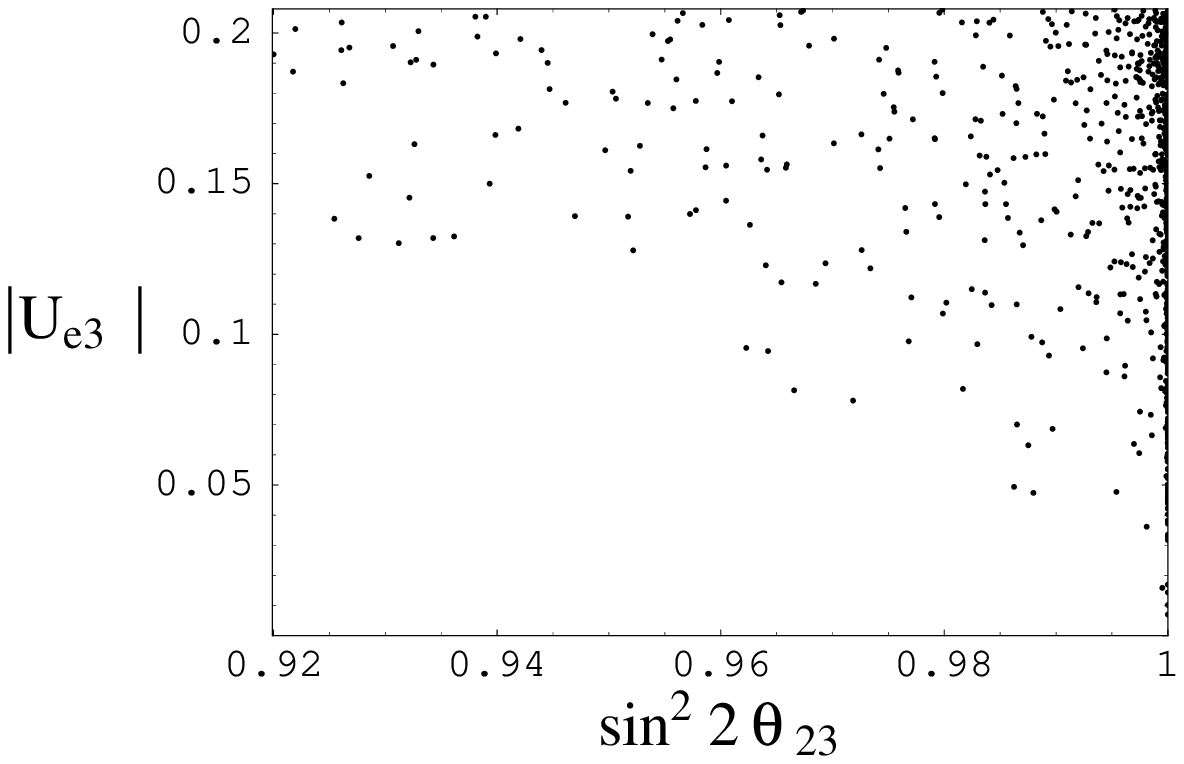}
\hspace{0.4cm}
\includegraphics[scale=.5]{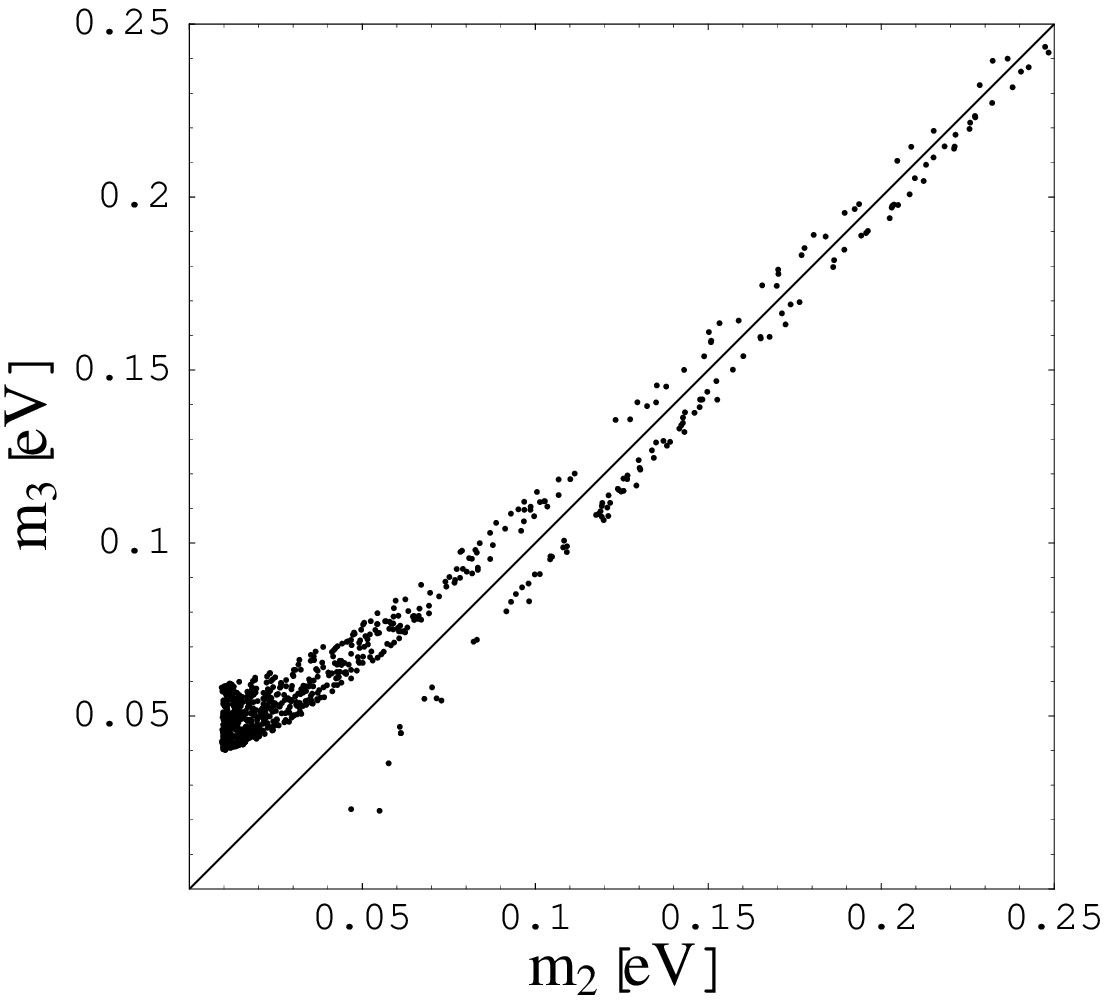}
\end{center}
\caption{Scatter plots of the allowed region in case 5,
where 100,000 random parameter sets are generated.}
\label{fig:c5}
\end{figure}

\begin{figure}
\begin{center}
\includegraphics[scale=.5]{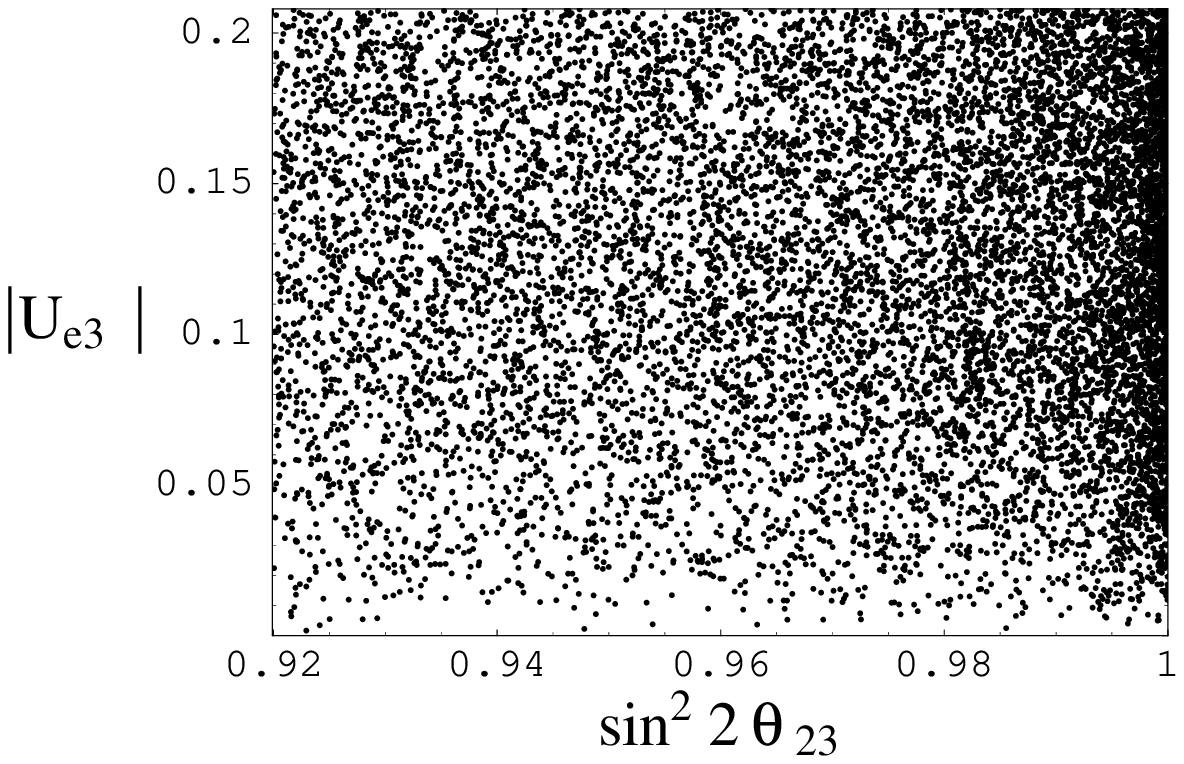}
\hspace{0.4cm}
\includegraphics[scale=.5]{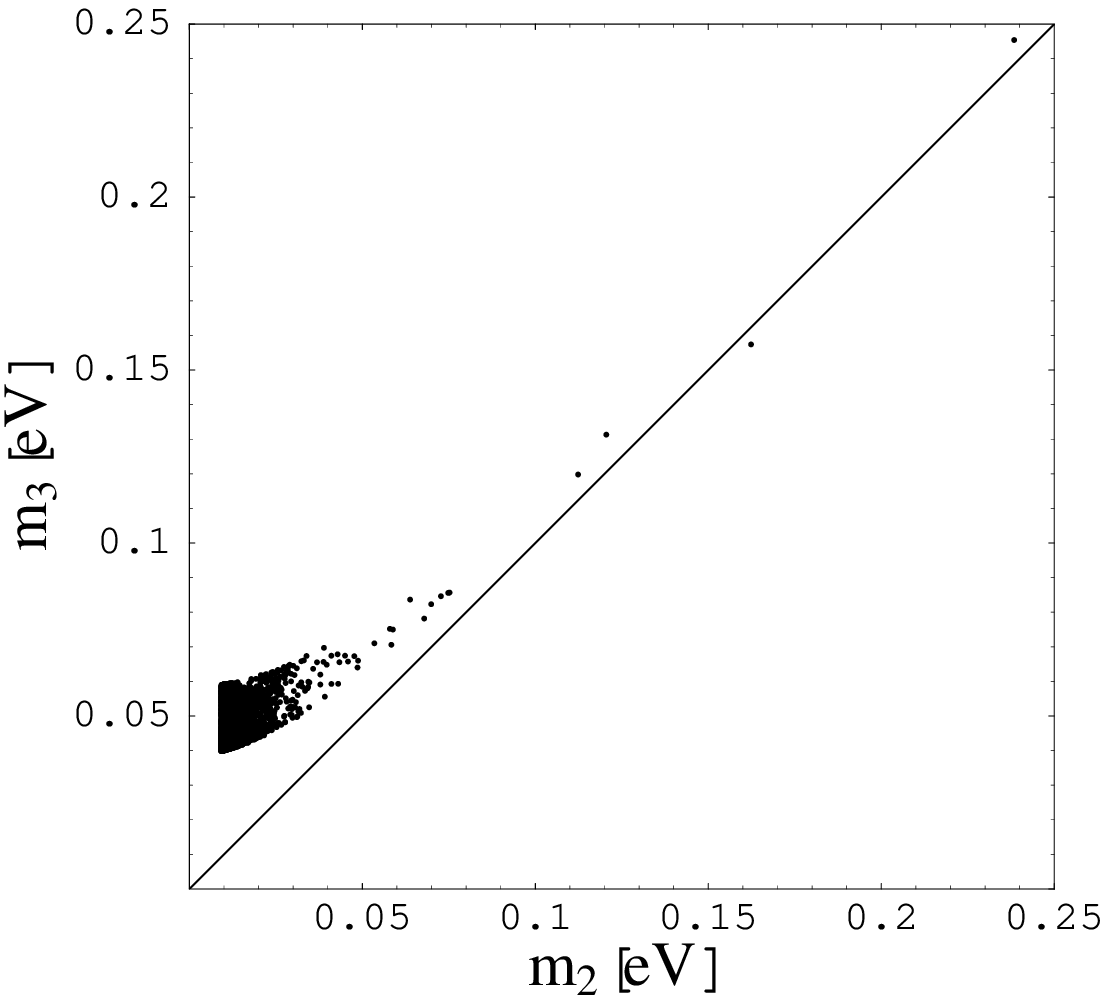}
\end{center}
\caption{Scatter plots of the allowed region in case 6,
where  100,000  random parameter sets are generated.}
\label{fig:c6}
\end{figure}

\begin{figure}
\begin{center}
\includegraphics[scale=.5]{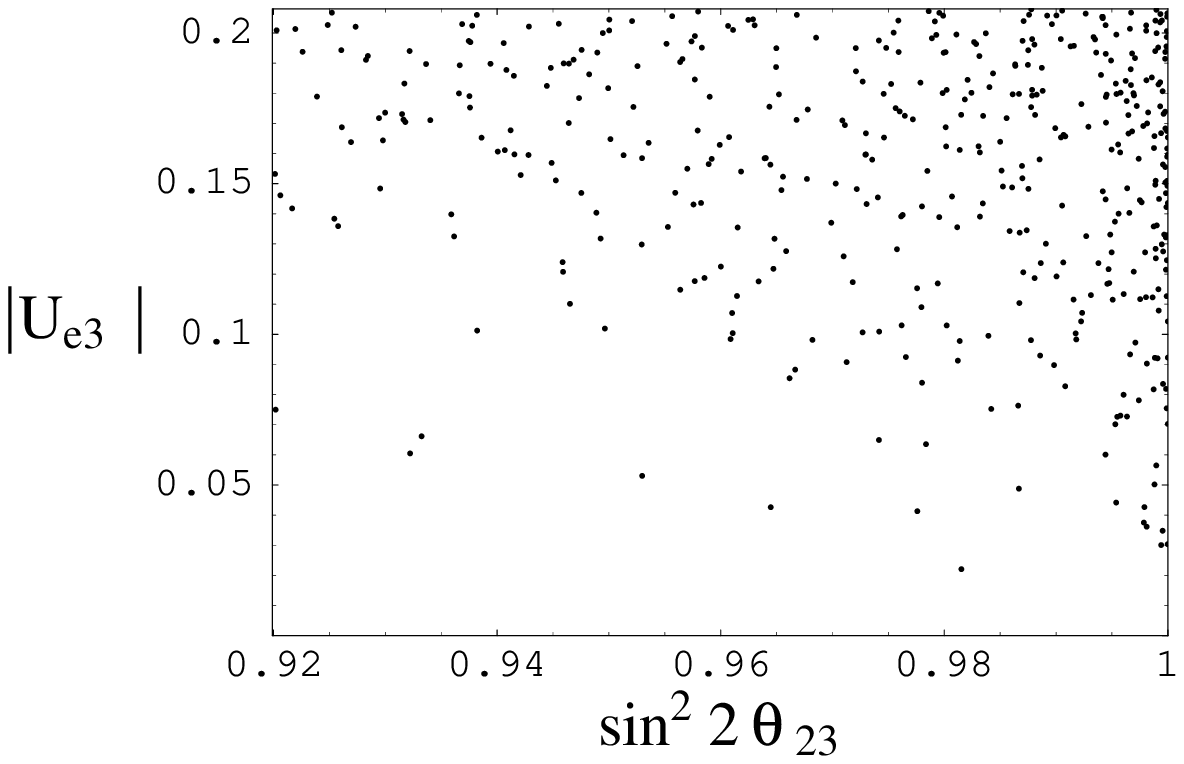}
\hspace{0.4cm}
\includegraphics[scale=.5]{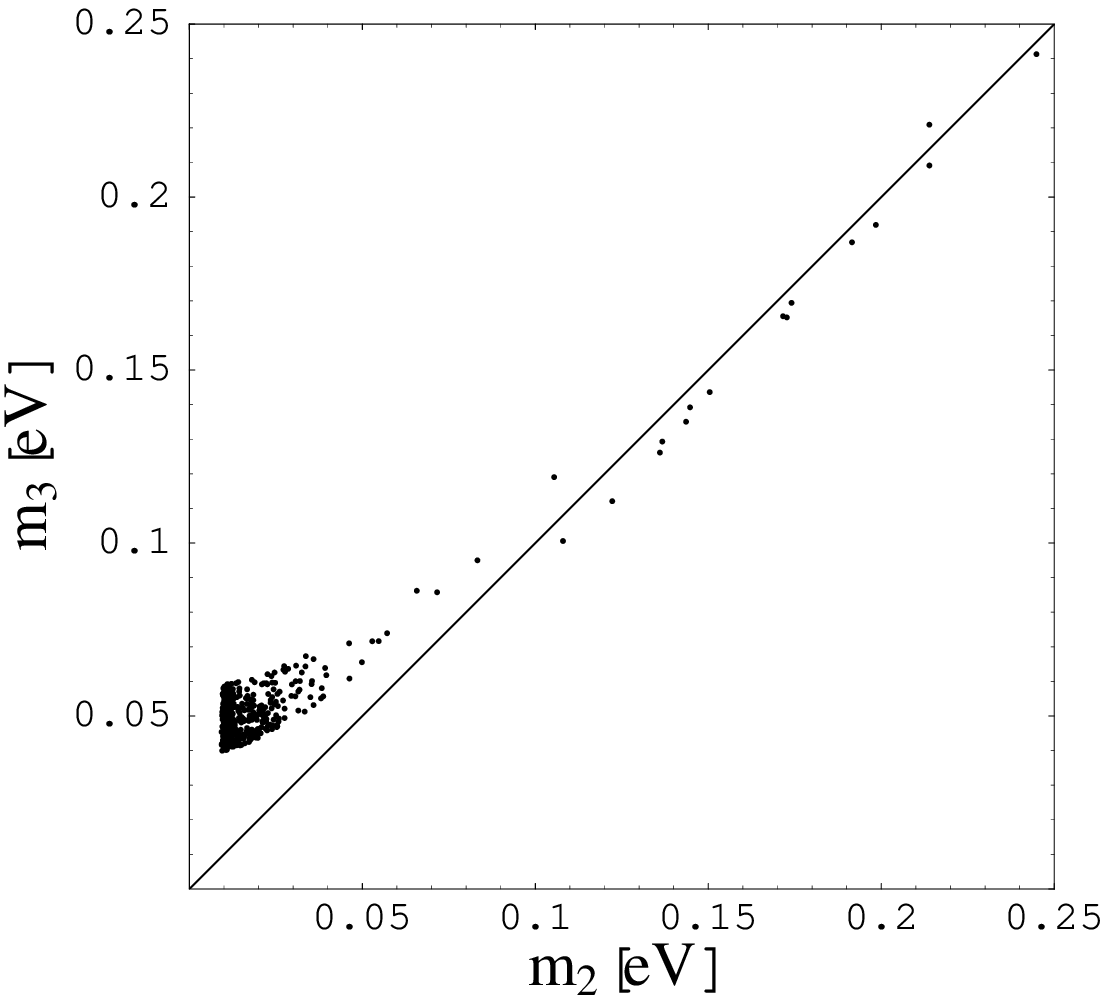}
\end{center}
\caption{Scatter plots of the allowed region in case 7,
where  100,000  random parameter sets are generated.}
\label{fig:c7}
\end{figure}

\begin{figure}
\begin{center}
\includegraphics[scale=.5]{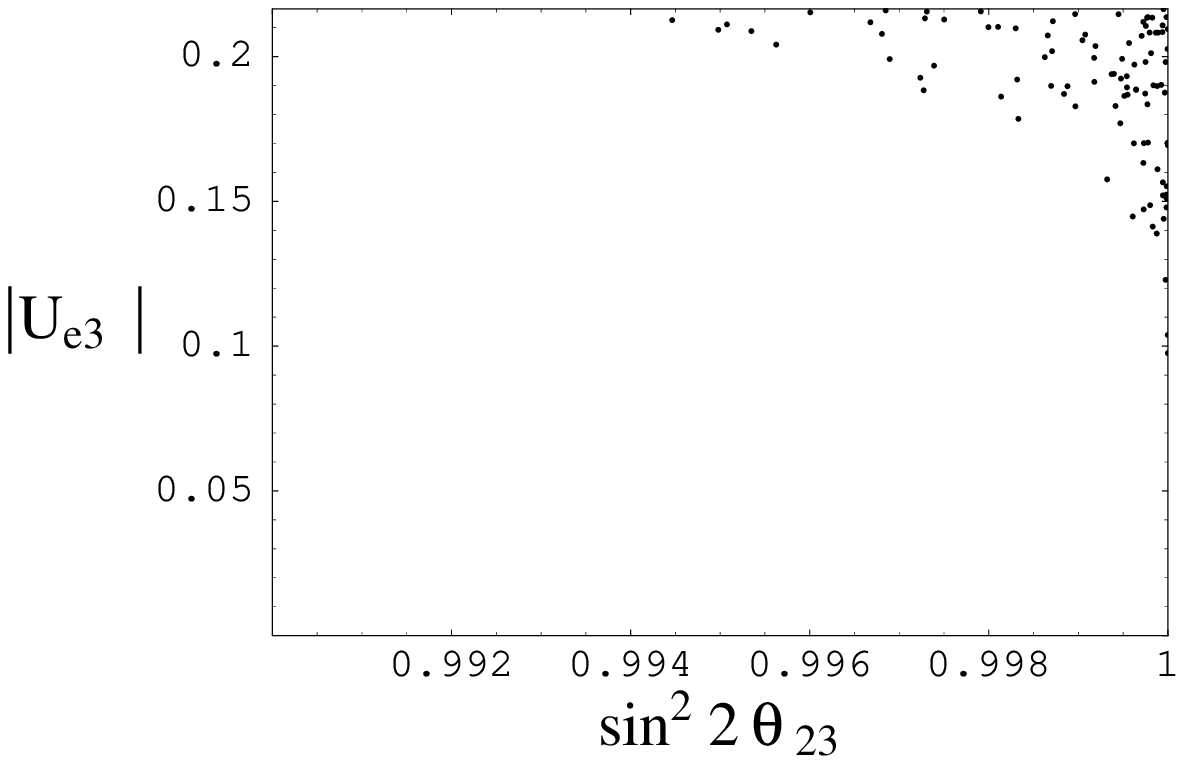}
\hspace{0.4cm}
\includegraphics[scale=.5]{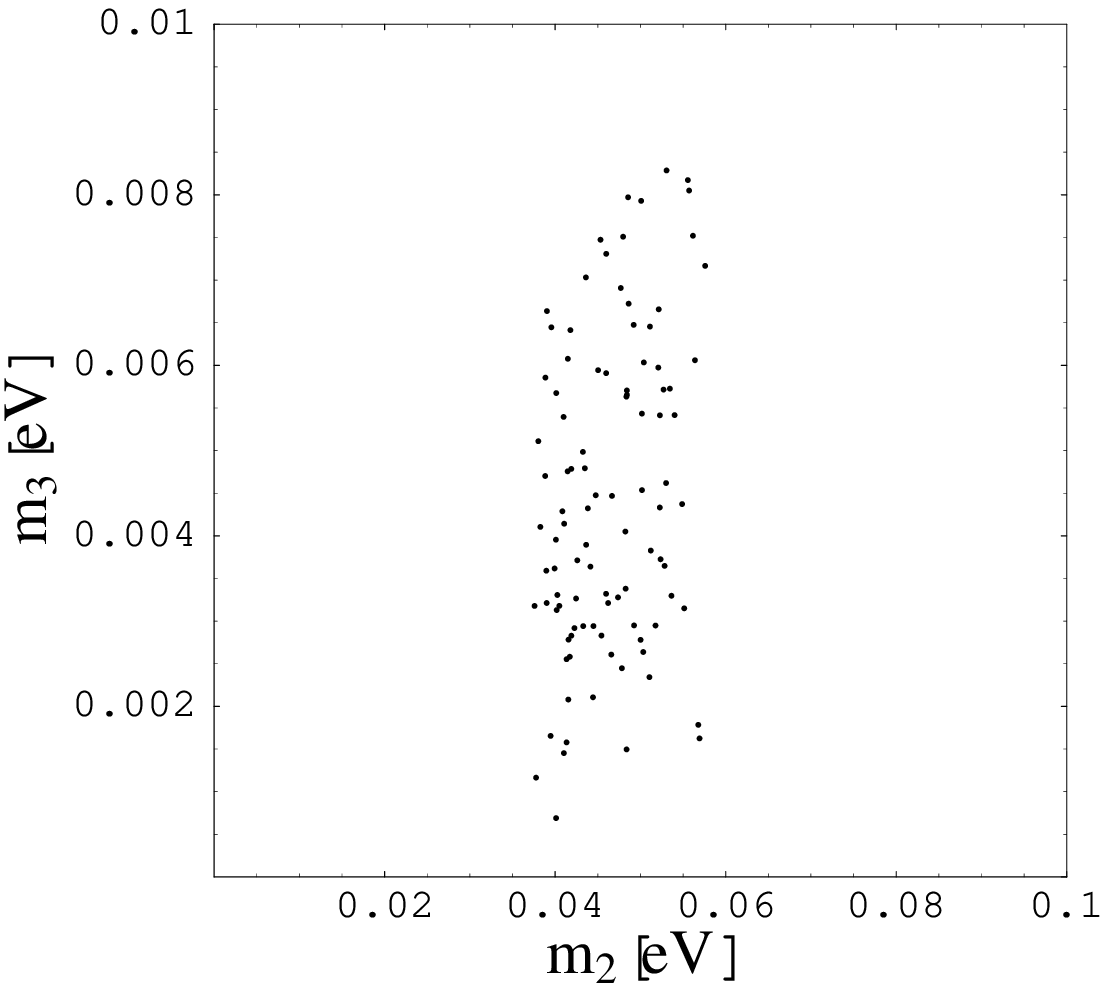}
\end{center}
\caption{Scatter plots of the allowed region in case 8,
where  2,000,000  random parameter sets are generated.}
\label{fig:c8}
\end{figure}

\begin{figure}
\begin{center}
\includegraphics[scale=.5]{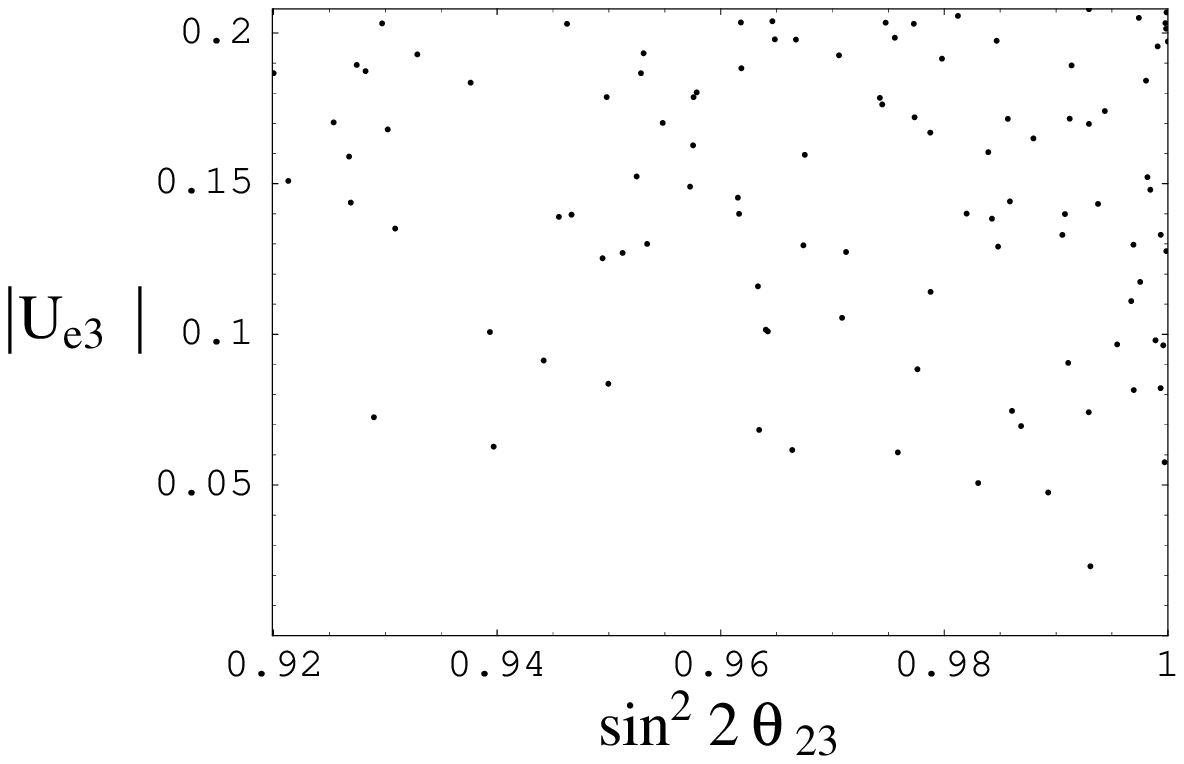}
\hspace{0.4cm}
\includegraphics[scale=.5]{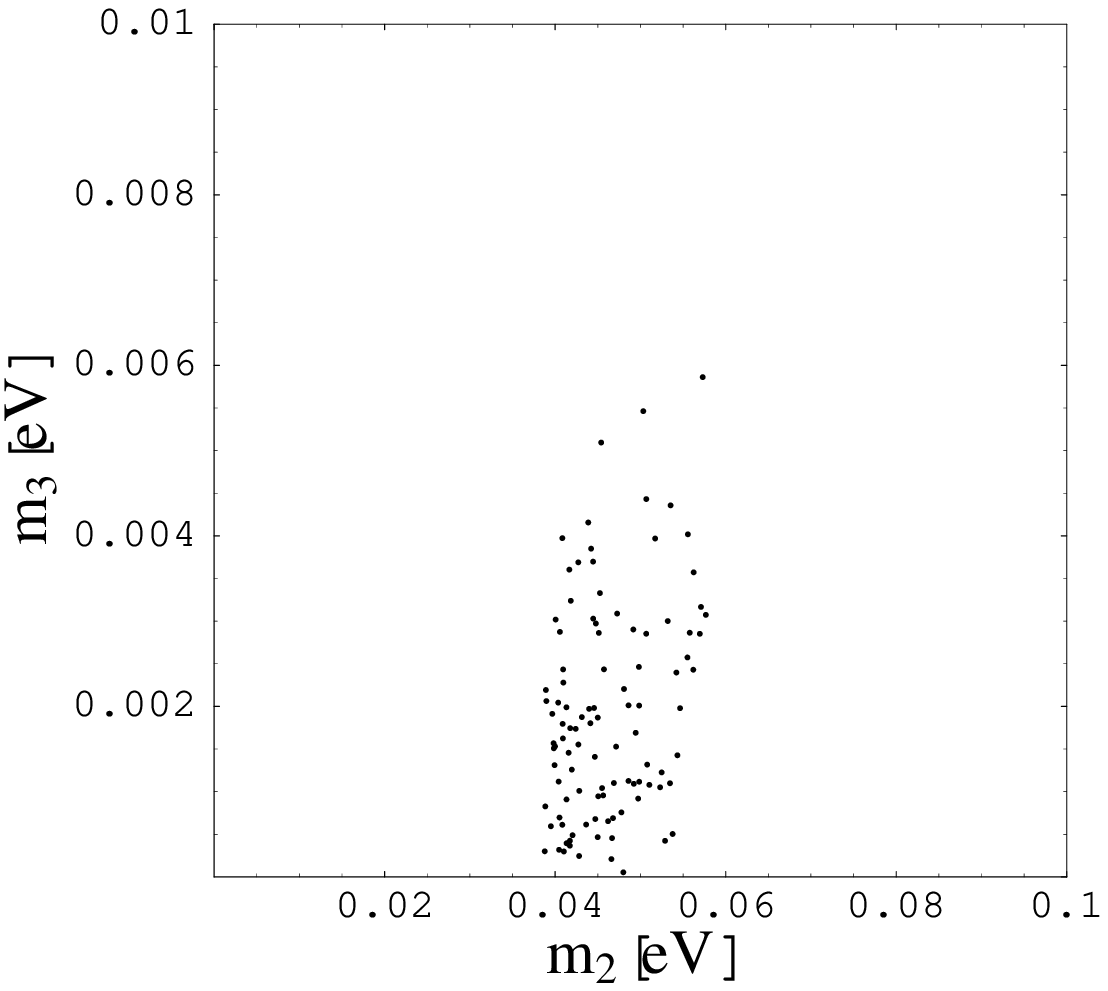}
\end{center}
\caption{Scatter plots of the allowed region in case 9,
where  100,000  random parameter sets are generated.}
\label{fig:c9}
\end{figure}

We have used a numerical approach to cases 5--9,
similar to the one in~\cite{hybrid}.
We have numerically generated random parameter sets
of the four observables $\theta_{12}$,
$\theta_{23}$,
$\theta_{13}$,
and $\delta$,
of the unphysical phase,
and of the mass $m_3$.
For each of these sets,
the neutrino mass ratios $m_1/m_3$ and $m_2/m_3$
have been computed by taking the absolute values
in the corresponding equations for $\mu_1/\mu_3$ and $\mu_2/\mu_3$;
by taking the absolute values of those equations
we avoid considering the two observables
on which we have hardly any experimental
grip---the Majorana phases $\Theta$ and $\Omega$.
Whenever the obtained neutrino mass ratios,
together with the inputted $m_3$,
were consistent with the experimental values for $m_2^2 - m_1^2$
and $\left| m_1^2 - m_3^2 \right|$,
we plotted the corresponding physical point in a scatter plot.
The resulting scatter plots of the allowed regions
for each model are presented in figures~\ref{fig:c5}--\ref{fig:c9}.
We have generated 100,000 random parameter sets
for each of the cases 5,
6,
7,
and 9;
for case 8 we have used 2,000,000 random parameter sets
because we have found that only a very small percentage
of them turned out to be allowed.

We present in figures~\ref{fig:c5}--\ref{fig:c9},
for each of cases~5--9,
respectively,
two types of scatter plots of the allowed regions:
in the $\sin^2{2 \theta_{23}}$--$\left| U_{e3} \right|$ plane,
and in the $m_2$--$m_3$ plane.

The neutrino mass spectrum in cases~5 and~7
may be of any possible type:
normal,
inverted,
or quasi-degenerate.
Case~6 typically displays a normal spectrum ($m_3 > m_2$).
Cases~8 and~9 have inverted neutrino mass spectra ($m_3 < m_2$),
as derived already in section~\ref{sec:case2} from the condition
$\left( \mnu^{-1} \right)_{ee} = 0$ alone.
Notice that case~8 has a distinct prediction:
it requires the atmospheric neutrino mixing
to be very close to maximal,
and $\left| U_{e3} \right|$ is also usually close to its upper bound;
this,
too,
we had already anticipated analytically in section~\ref{sec:case2}.

For all cases~5--9,
$\left| U_{e3} \right|$ cannot vanish,
but it may be very small.
We have performed a dedicated search
of points with very small $\left| U_{e3} \right|$
by constructing random parameter sets with very small inputted $\theta_{13}$.
We thereby found the lower bounds on $\left| U_{e3} \right|$
given in table~\ref{tab:massrate}.
\begin{table}[h]
\begin{center}
\begin{tabular}{|c|c|c|c|c|c|} \hline
 & case 5 & case 6 & case 7 & case 8 &  case 9\\ \hline
$|U_{e3}| > $ & $0.001$ & $10^{-6}$ &$0.0005$ & $0.04$ & $0.0003$\\ \hline
$\langle m \rangle_{ee}\, \mathrm{(eV)} >$ &
$0.0001$ & $10^{-5}$ & $0.0001$ & $0.01$ & $0.01$\\  \hline
\end{tabular} 
\caption{Lower bounds of $\left| U_{e3} \right|$
and on $\langle m \rangle_{ee}$,
for each case.}
\label{tab:massrate}
\end{center}
\end{table} 
One sees that,
in all cases but in case~8,
$\left| U_{e3} \right|$ can attain values so low
as to be experimentally undistinguishable from zero.

\begin{figure}
\begin{center}
\includegraphics[scale=.43]{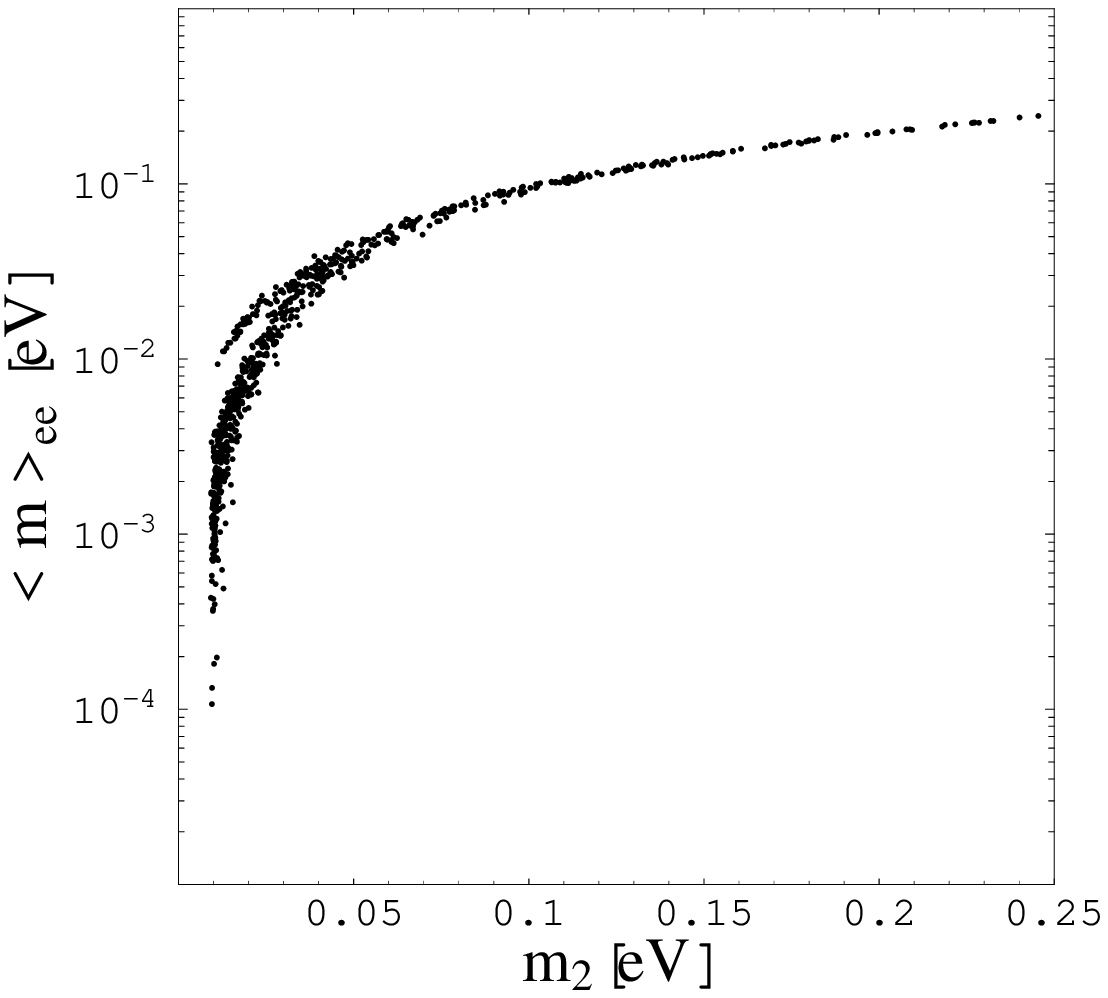}
\hspace{0.4cm}
\includegraphics[scale=.43]{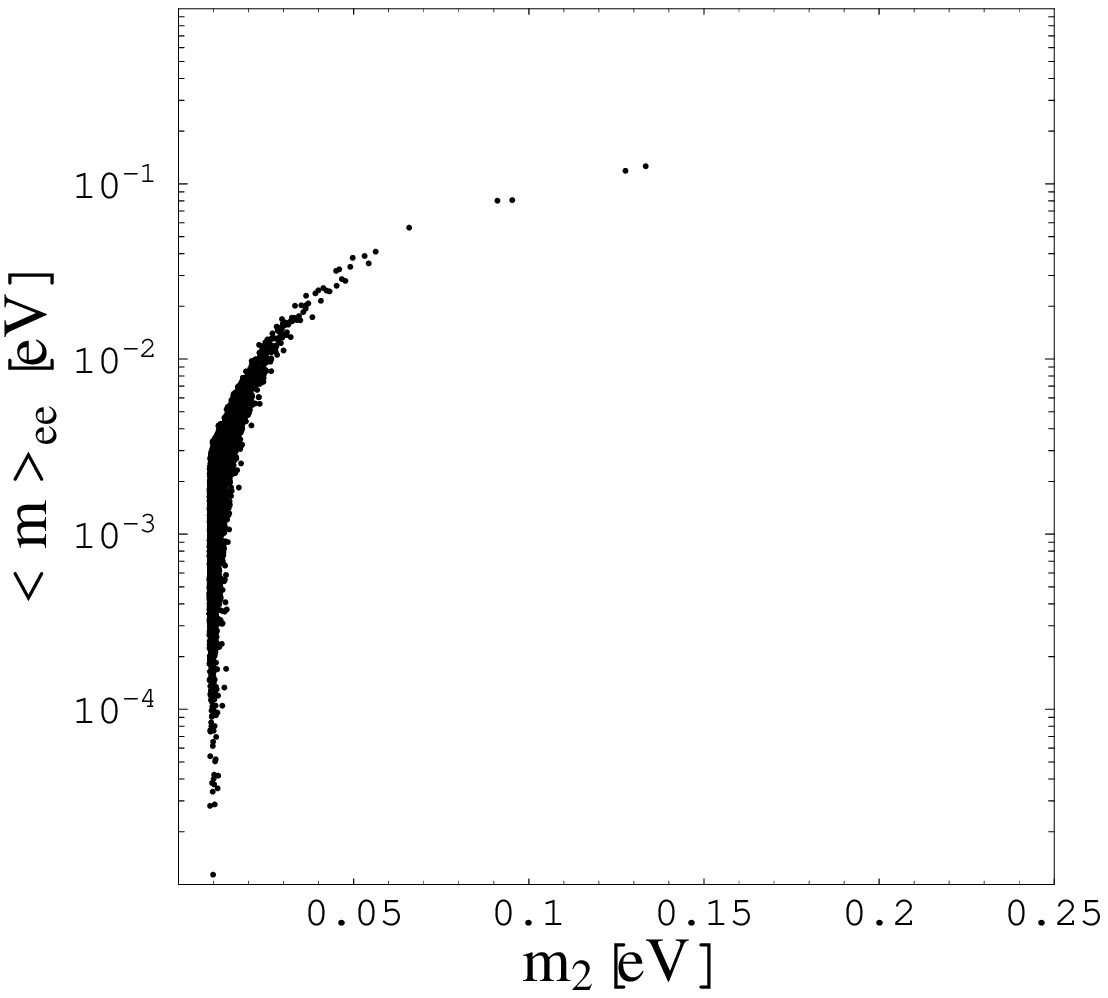}
\hspace{0.4cm}
\includegraphics[scale=.43]{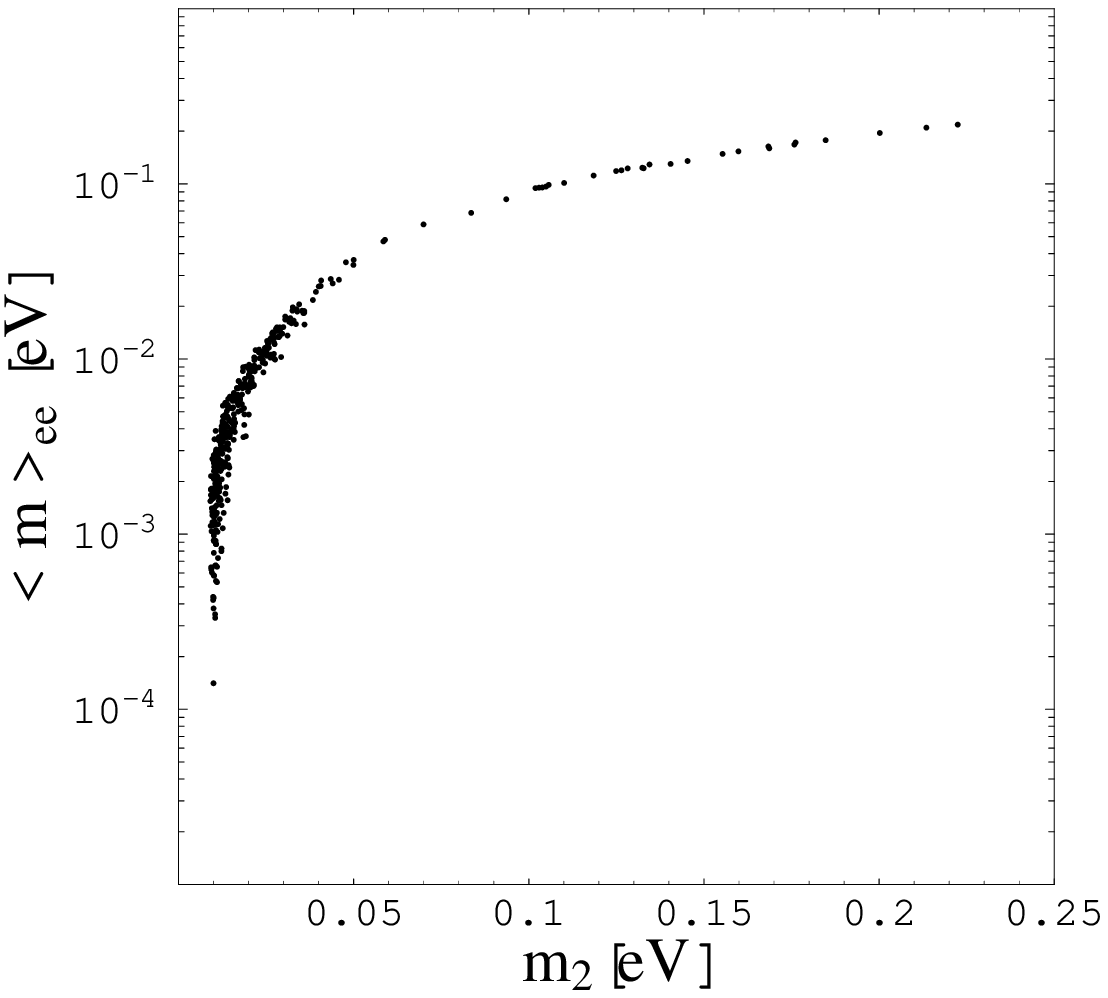}
\end{center}
\caption{Scatter plots of the $\langle m \rangle_{ee}$
of neutrinoless $\beta\beta$ decay as a function of $m_2$ for cases~5,
6,
and~7.}
\label{fig:beta567}
\end{figure}

\begin{figure}
\begin{center}
\includegraphics[scale=.63]{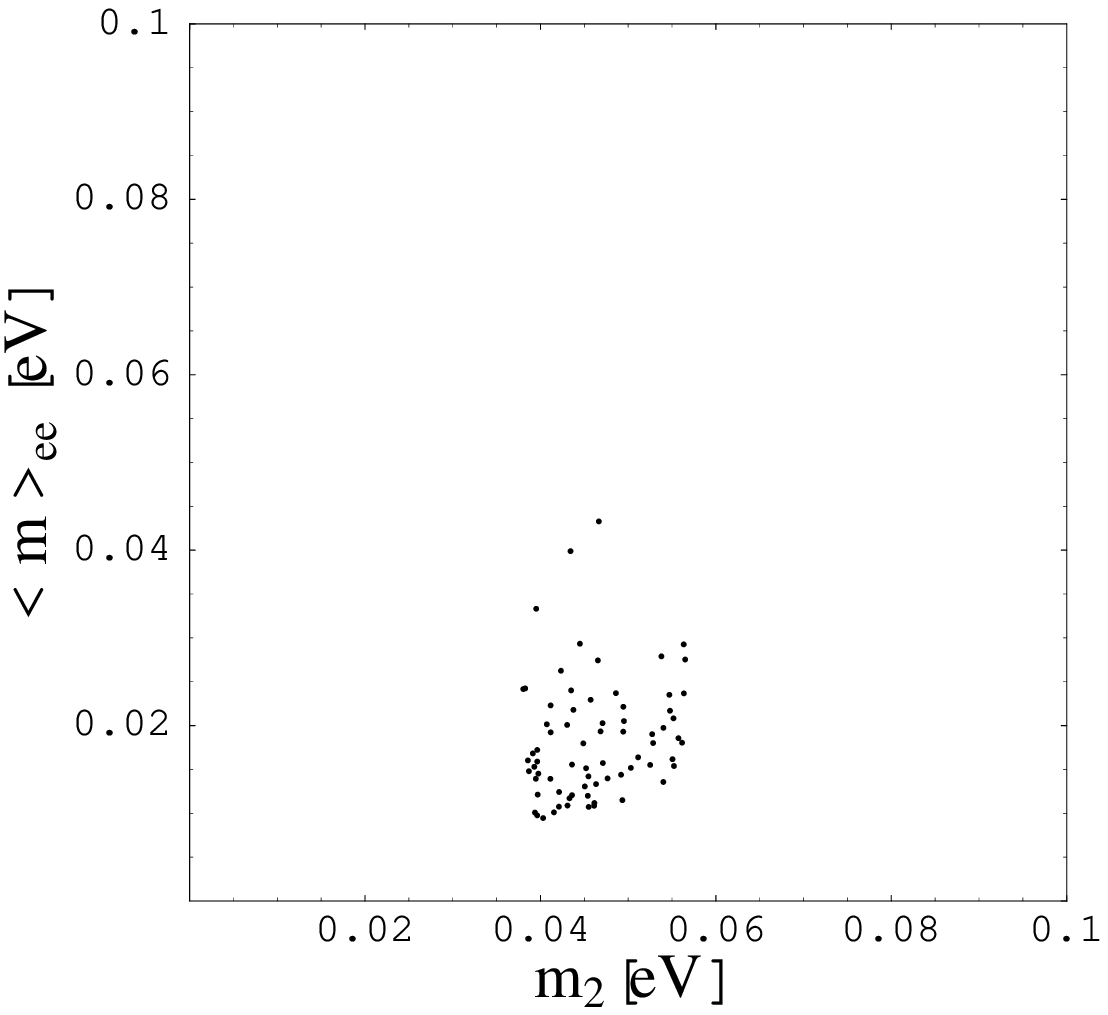}
\hspace{0.4cm}
\includegraphics[scale=.63]{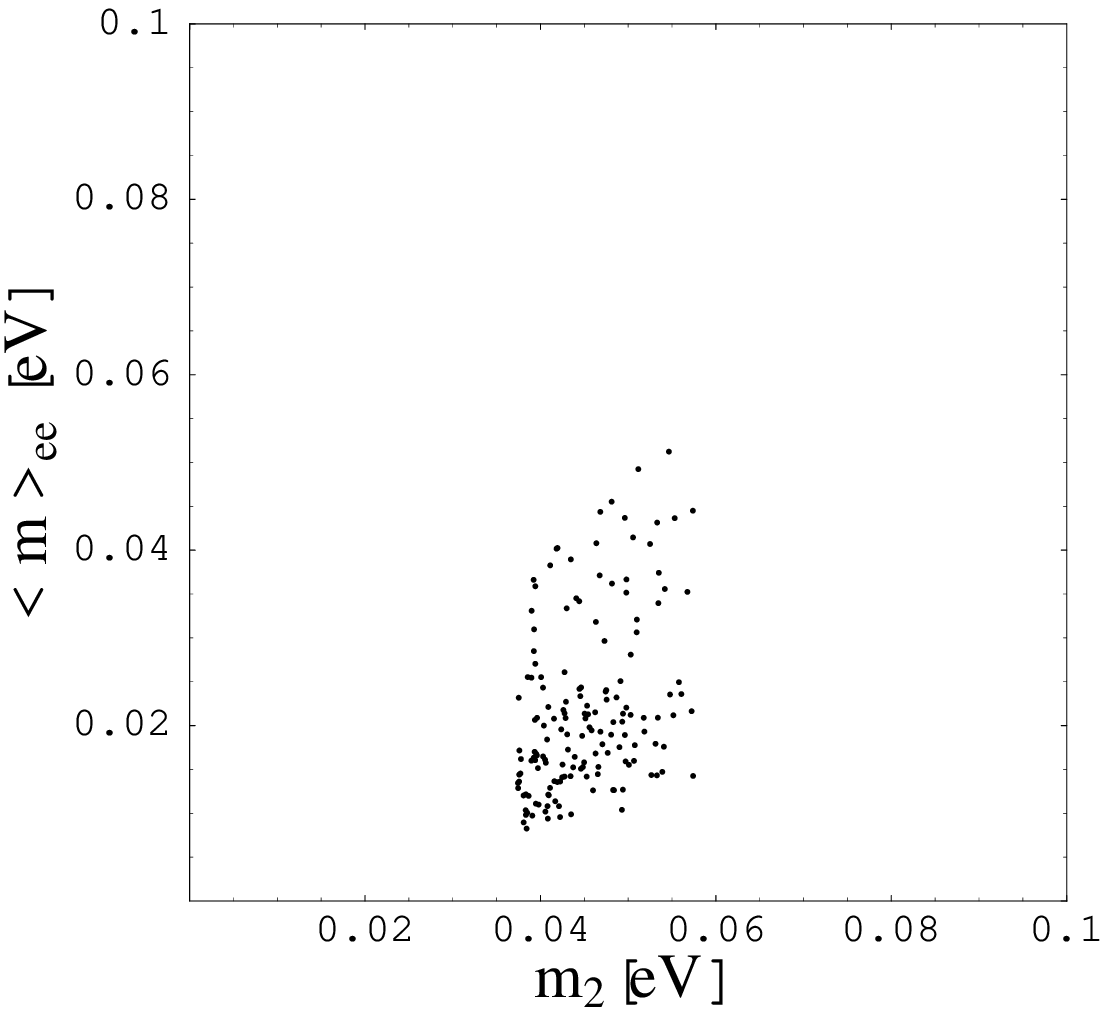}
\end{center}
\caption{Scatter plots of the $\langle m \rangle_{ee}$
of neutrinoless $\beta\beta$ decay as a function of $m_2$
for cases~8 and~9.}
\label{fig:beta89}
\end{figure}

We have also investigated the decay rate
for neutrinoless double-beta decay in each model. 
That rate is controlled by the effective Majorana mass
\be
\langle m \rangle_{ee}=\left|
m_1 c_{13}^2 c_{12}^2 e^{i \Theta}
+ m_2 c_{13}^2 s_{12}^2
+ m_3 s_{13}^2 e^{i \left( \Omega - 2 \delta \right)} 
\right|.
\ee
The values of the Majorana phases are extracted from the equations
for $\mu_1 / \mu_3$ and $\mu_2 / \mu_3$.
We present scatter plots of $\langle m \rangle_{ee}$
as a function of $m_2$,
for each case,
in figures~\ref{fig:beta567} and~\ref{fig:beta89}.
One sees in those figures that $\langle m \rangle_{ee}$
is at most 0.3~eV in cases~5--7;
in cases~8 and~9 the upper bound is about an order of magnitude smaller.
Lower bounds of  $\langle m \rangle_{ee}$
are given in table~\ref{tab:massrate};
since cases~8 and~9 have an inverted neutrino mass spectrum,
the lower bound on $\langle m \rangle_{ee}$
is not so suppressed there as in cases~5--7.

\section{Summary}

The type-I seesaw mechanism has three sources of lepton mixing:
the charged-lepton mass matrix $M_\ell$,
the neutrino Dirac mass matrix $M_D$,
and the mass matrix $M_R$ of the right-handed neutrinos.
It is interesting to choose $M_R$
as the only source of mixing.
A symmetry reason for this choice is provided
by the assumption of conservation of the family lepton numbers
$L_\alpha$ ($\alpha = e,\,\mu,\,\tau$)
in all terms of dimension 4 in the Lagrangian,
because in that case $M_\ell$ and $M_D$
are automatically diagonal;
the Majorana mass terms of the right-handed neutrinos
break the $L_\alpha$ softly
and we are therefore allowed to obtain lepton mixing
from $M_R$~\cite{Z2model}.
This framework admits an arbitrary number of Higgs doublets,
as their Yukawa-coupling matrices are all diagonal,
hence flavour-changing Yukawa neutral interactions
among the charged leptons are strongly suppressed~\cite{loop}.
In this framework,
it looks sensible to consider relations
among the matrix elements of the \emph{inverted}
neutrino mass matrix $\mnu^{-1}$~\cite{lavoura}. 

In this paper,
we have used as a starting point
a $\mu$--$\tau$ interchange \emph{anti}symmetry in $\mnu^{-1}$.
We have shown that this can be obtained
by imposing on the full Lagrangian
a symmetry of the $\mathbbm{Z}_4$ type.
However,
such a $\mu$--$\tau$ antisymmetric matrix $\mnu^{-1}$ is singular,
therefore $\mu$--$\tau$ symmetric ``perturbations''
must be added to it.
We have done so by introducing complex scalar
$SU(2) \times U(1)$ invariants
with Yukawa couplings to the right-handed neutrinos.
These scalar fields have non-trivial lepton numbers $L_\alpha$,
and there are four basic possibilities to introduce such singlets
in accord with the above-mentioned $\mathbbm{Z}_4$---see section~2.
The VEVs of these scalar $SU(2) \times U(1)$ invariants
generate the $\mu$--$\tau$ symmetric ``perturbations'' in $\mnu^{-1}$. 

Each of the four basic possibilities
leads to a four-parameter neutrino mass matrix
(cases~1, 2, 3, and~4 in section~2)
which we have shown to be ruled out by the data. 
As a next step,
we have considered the six combinations
of the four basic possibilities.  
One of them leads to a five-parameter mass matrix
(case~10)
and is ruled out as well.
On the other hand,
the remaining cases~5, 6, 7, 8, and~9,
which have neutrino mass matrices with six parameters, 
are physically viable,
as we have shown by numerical analysis.
We have refrained from combining
more than two of the four basic possibilities,
since the ensuing neutrino mass matrices
presumably have negligible predictive power.

The five viable cases have in common that they allow $s_{13}^2$
to be as large as the present upper bound.
The numerical scan in the parameter space of cases~5,
6,
and~7 shows a preference for a hierarchical spectrum,
although quasi-degenerate or inverted spectra cannot be ruled out.
The neutrino mass spectrum in cases 8 and 9 has inverted hierarchy;
this follows solely from the condition
$\left( \mnu^{-1} \right)_{ee} = 0$,
which has been analytically analyzed in section~5,
an analysis which is borne out by our numerical results---see
figures~4 and~5.
The most predictive neutrino mass matrix is the one of case 8,
because it leads to practically maximal atmospheric mixing,
correlated with a large $s_{13}^2$---see
the analytical discussion of conditions~(\ref{one})
and~(\ref{three}) in section~5,
and the numerical result in figure~4.

The research presented in this paper displays once again the usefulness
of conditions on the matrix elements
of the \emph{inverted} neutrino mass matrix $\mnu^{-1}$. 
Indeed,
in the framework employed here,
it is most natural that symmetries imposed on the Lagrangian
lead to relations in $\mnu^{-1}$.
In this paper
we have used $\mu$--$\tau$ interchange \emph{antisymmetry},
instead of the well-known $\mu$--$\tau$ interchange symmetry,
as the basic ingredient to obtain viable
and predictive neutrino mass matrices.

\vspace*{5mm}

\paragraph{Acknowledgements}
The work of S.K.\ was supported by the Japanese Society
for the Promotion of Science.
The work of L.L.\ was supported by the Portuguese
\textit{Funda\c c\~ao para a Ci\^encia e a Tecnologia}
through the projects POCTI/FNU/44409/2002
and U777--Plurianual.
The work of M.T.\ is supported by the
Grant-in-Aid for Science Research
of the Ministry of Education, Science, and Culture of Japan
No.\ 16028205, No.\ 17540243.

\clearpage


\newpage

\begin{thebibliography}{99}

\bibitem{tortola}
M.~Maltoni, T.~Schwetz, M.~Tortola and J.W.F.~Valle,
\textit{Status of global fits to neutrino oscillations},
\textit{New J.~Phys.}~\textbf{6} (2004) 122 [hep-ph/0405172].

\bibitem{fogli}
G.L.~Fogli, E.~Lisi, A.~Marrone and A.~Palazzo,
\textit{Global analysis of three-flavor neutrino masses and mixings},
hep-ph/0506083.

\bibitem{altarelli}
G.~Altarelli and F.~Feruglio, 
\textit{Models of neutrino masses and mixings},
\textit{New J. Phys.}~\textbf{6} (2004) 106 [hep-ph/0405048]; \\
G.~Altarelli,
\textit{Normal and special models of neutrino masses and mixings},
to be published in the
\textit{Proceedings of 19th Rencontres de physique de la vall\'ee d'Aoste:
Results and perspectives in particle physics,
La Thuile, Aosta valley, Italy, 27~Feb.--5~Mar.~2005}
[hep-ph/0508053]; \\
R.N.~Mohapatra et al.,
\textit{Theory of neutrinos: A white paper},
hep-ph/0510213.

\bibitem{early}
T.~Fukuyama and H.~Nishiura, 
\textit{Mass matrix of Majorana neutrinos}, 
hep-ph/9702253; \\
E.~Ma and M.~Raidal, 
\textit{Neutrino mass, muon anomalous magnetic moment,
and lepton flavor nonconservation},
\textit{Phys. Rev. Lett.} \textbf{87} (2001) 011802
[hep-ph/0102255]; 
\textit{Err. ibid.} \textbf{87} (2001) 159901; \\
C.S.~Lam, 
\textit{A 2-3 symmetry in neutrino oscillations},
\textit{Phys. Lett.} \textbf{B~507} (2001) 214 
[hep-ph/0104116]; \\
K.R.S.~Balaji, W.~Grimus and T.~Schwetz,
\textit{The solar LMA neutrino oscillation solution
in the Zee model},
\textit{Phys. Lett.} \textbf{B~508} (2001) 301
[hep-ph/0104035]; \\
E.~Ma,
\textit{The all-purpose neutrino mass matrix},
\textit{Phys. Rev.} \textbf{D~66} (2002) 117301
[hep-ph/0207352]; \\
P.F.~Harrison and W.G.~Scott, 
\textit{$\mu - \tau$ reflection symmetry in lepton mixing
and neutrino oscillations}, 
\textit{Phys. Lett.} \textbf{B~547} (2002) 219
[hep-ph/0210197].

\bibitem{joshipura}
For specific schemes and additional references,
see for instance \\
Y.~Koide, 
\textit{Universal texture of quark and lepton mass matrices
with an extended flavor $2 \leftrightarrow 3$ symmetry},
\textit{Phys.~Rev.}~\textbf{D~69} (2004) 093001 
[hep-ph/0312207]; \\
W.~Grimus, A.S.~Joshipura, S.~Kaneko,
L.~Lavoura, H.~Sawanaka and M.~Tanimoto, 
\textit{Non-vanishing $U_{e3}$ and $\cos{2 \theta_{23}}$
from a broken $Z_2$ symmetry},
\textit{Nucl.~Phys.}~\textbf{B~713} (2005) 151 [hep-ph/0408123]; \\
W.~Grimus and L.~Lavoura,
\textit{$S_3 \times \mathbbm{Z}_2$ model for neutrino mass matrices},
\textit{JHEP} \textbf{08} (2005) 013 [hep-ph/0504153]; \\
R.N.~Mohapatra and W.~Rodejohann, 
\textit{Broken $\mu-\tau$ symmetry
and leptonic $CP$ violation},
\textit{Phys.~Rev.} \textbf{D~72} (2005) 053001 [hep-ph/0507312].

\bibitem{aizawa}
T. Kitabayashi and M. Yasue,
\textit{$\mu-\tau$ symmetry and maximal CP violation},
\textit{Phys. Lett.}~\textbf{B~621} (2005) 133
[hep-ph/0504212]; \\
I. Aizawa, T. Kitabayashi and M. Yasue,
\textit{Determination of neutrino mass texture for maximal CP violation},
\textit{Nucl. Phys.} \textbf{B~728} (2005) 220 
[hep-ph/0507332]; \\
T. Kitabayashi and M. Yasue,
\textit{A new type of complex neutrino mass texture and $\mu-\tau$ symmetry},
hep-ph/0510132.

\bibitem{Z2model}
W.~Grimus and L.~Lavoura,
\textit{Softly broken lepton numbers and maximal neutrino mixing},
\textit{JHEP} \textbf{07} (2001) 045 [hep-ph/0105212].

\bibitem{seesaw}
P.~Minkowski,
\textit{$\mu \to e \gamma$ at a rate of one out of $10^9$ muon decays?},
\textit{Phys.~Lett.}~\textbf{B~67} (1977) 421; \\
T.~Yanagida,
in \textit{Proceedings of the workshop on unified theory
and baryon number in the universe},
O.~Sawata and A.~Sugamoto eds.,
KEK report \textbf{79-18},
Tsukuba, Japan 1979; \\
S.L.~Glashow,
in \textit{Quarks and leptons,
proceedings of the advanced study institute
(Carg\`ese, Corsica, 1979)},
J.-L.~Basdevant et al.~eds.,
Plenum, New York 1981; \\
M.~Gell-Mann, P.~Ramond and R.~Slansky,
\textit{Complex spinors and unified theories},
in \textit{Supergravity},
D.Z.~Freedman and F.~van~Nieuwenhuizen eds.,
North Holland, Amsterdam 1979; \\
R.N.~Mohapatra and G.~Senjanovic,
\textit{Neutrino mass and spontaneous parity violation},
\textit{Phys.~Rev.~Lett.}~\textbf{44} (1980) 912.

\bibitem{seesaw1}
For detailed discussions of the seesaw mechanism, see \\
J.~Schechter and J.W.F.~Valle,
\textit{Neutrino masses in $SU(2) \times U(1)$ theories},
\textit{Phys. Rev.}~\textbf{D~22} (1980) 2227; \\
S.M.~Bilenky, J.~Hosek and S.T.~Petcov,
\textit{On oscillations of neutrinos with Dirac and Majorana masses},
\textit{Phys.~Lett.} \textbf{94B} (1980) 495; \\
I.Yu.~Kobzarev, B.V.~Martemyanov, L.B.~Okun and M.G.~Shchepkin, 
\textit{The phenomenology of neutrino oscillations},
\textit{Yad.~Phys.} \textbf{32} (1980) 1590 
[\textit{Sov.~J.~Nucl. Phys.}~\textbf{32} (1981) 823]; \\
J.~Schechter and J.W.F.~Valle,
\textit{Neutrino decay and spontaneous violation of lepton number},
\textit{Phys.~Rev.} \textbf{D~25} (1982) 774; \\
W.~Grimus and L.~Lavoura,
\textit{The seesaw mechanism at arbitrary order: disentangling the
small scale from the large scale},
\textit{JHEP} \textbf{11} (2000) 042 [hep-ph/0008179].

\bibitem{lavoura}
L.~Lavoura, 
\textit{Zeros of the inverted neutrino mass matrix},
\textit{Phys.~Lett.}~\textbf{B~609} (2005) 317 [hep-ph/0411232].

\bibitem{fritzsch}
H.~Fritzsch,
\textit{Weak interaction mixing in the six-quark theory},
\textit{Phys.~Lett.}~\textbf{73B} (1978) 317; \\
L.-F.~Li,
\textit{Comments on the derivation of the mixing angles},
\textit{Phys.~Lett.}~\textbf{84B} (1979) 461.

\bibitem{hybrid}
S.~Kaneko, H.~Sawanaka and M.~Tanimoto,
\textit{Hybrid textures of neutrinos},
\textit{JHEP} \textbf{08} (2005) 073 [hep-ph/0504074].

\bibitem{loop}
W.~Grimus and L.~Lavoura,
\textit{Soft lepton-flavor violation
in a multi-Higgs-doublet seesaw model},
\textit{Phys.~Rev.} \textbf{D~66} (2002) 014016 [hep-ph/0204070].

\end{thebibliography}
\end{document}